\documentclass[a4paper, 11pt]{article}

\usepackage[a4paper, margin=2.5cm]{geometry}
\usepackage[latin1]{inputenc}
\usepackage[francais,english]{babel}
\usepackage{amsmath,amssymb}
\usepackage{datetime,mathdots}
\usepackage{hyperref}\hypersetup{colorlinks=true}
\usepackage{graphicx}
\usepackage[percent]{overpic}
\usepackage{setspace}
\usepackage{anyfontsize}

\def\be{\begin{equation}}
\def\ee{\end{equation}}
\def\bea{\begin{eqnarray}}
\def\eea{\end{eqnarray}}
\def\bal{\begin{align}}
\def\eal{\end{align}}

\begin{document} 

\begin{center}
{\Large\bf Cultural Structures of Knowledge \\
from Wikipedia Networks of First Links}

\vspace{0.2cm}

{\bf Maxime Gabella}

\vspace{0.1cm}

{\it Institute for Advanced Study, Einstein Drive, Princeton, New Jersey 08540, USA}
\end{center}

\vspace{0.3cm}

{\abstract
Knowledge is useless without structure. 
While the classification of knowledge has been an enduring philosophical enterprise, it recently found applications in computer science, notably for artificial intelligence.
The availability of large databases allowed for complex ontologies to be built automatically, for example by extracting structured content from Wikipedia.
However, this approach is subject to manual categorization decisions made by online editors.
Here we show that an implicit classification hierarchy emerges spontaneously on Wikipedia. 
We study the network of first links between articles, and find that it centers on a core cycle involving concepts of fundamental classifying importance.
We argue that this structure is rooted in cultural history.
For European languages, articles like Philosophy and Science are central, whereas Human and Earth dominate for East Asian languages. This reflects the differences between ancient Greek thought and Chinese tradition. Our results reveal the powerful influence of culture on the intrinsic architecture of complex data sets.}

%%%%%%%%%%%%%%%%%%%%%%%%%%%%%%
\section{Introduction}

The fact that anyone is free to edit articles on Wikipedia could have resulted in a disaster. Instead, articles are often of high quality, with a level of accuracy similar to the reputable Encyclop\ae dia Britannica, as demonstrated by a 2005 Nature investigation \cite{WikiBrit}. Moreover, in the absence of global coordination among editors, no definite large-scale structure was expected. Yet, in 2008 some universal property was discovered \cite{WikiPhilo} (see also~\cite{IK}): clicking on the first link in the main text of any article and  repeating the process for subsequent articles almost always led to the Philosophy article.
An explanation of this surprising phenomenon could emanate from the classifying nature of first links. Wikipedia's Manual of Style~\cite{WikiStyle} indeed recommends that ``the opening sentence should provide links to the broader or more elementary topics that are important to the article's topic or place it into the context where it is notable.'' This implies that following the sequence of first links would naturally lead to increasingly fundamental concepts. 
In turn, the prominence of Philosophy in this spontaneous classification could then stem from the foundational role played by ancient Greek philosophy in the intellectual development of Western civilization.
It would thus appear that the self-organization of knowledge on Wikipedia is informed by cultural heritage. We will test this hypothesis by studying the network of first links on Wikipedia. 

Many studies have analyzed Wikipedia as a complex network, with articles represented by vertices and links by edges. 
The semantic structure of the English edition of Wikipedia was analyzed in~\cite{DBLP:journals/corr/abs-cs-0512085}.
References~\cite{PhysRevE.74.036116} and~\cite{PhysRevE.74.016115} found that Wikipedia networks in several languages were ``scale-free''~\cite{Barabasi412}\cite{RePEc:oxp:obooks:9780199665174} and related their growths to the ``preferential attachement'' mechanism~\cite{1999Sci...286..509B}.
Properties of Wikipedia networks were further studied in~\cite{2008EL.....8128006C}\cite{2011PLoSO...617333M}\cite{2011EL.....9358005Z}.

In contrast to these works, which usually considered (a subset of) all the links on Wikipedia,  we focus on the network made out only of first links, because of their classifying function.
The network of first links for the English Wikipedia has already been found to be scale-free in~\cite{2016arXiv160500309I}.
Here we argue that culture influences the architecture of first-link networks for Wikipedias in different languages.

%%%%%%%%%%%%%%%%%%%%%%%%%%%%%%
\section{Core cycle of Wikipedia}

By first link of a Wikipedia article we mean more precisely the first link in the main text (not in parentheses), and exclude links in comments, tables, or images, as well as self-links and links to inexistent or external pages.
Each vertex has exactly one outgoing edge (its first link), but can have many incoming edges (when it is the first link of many articles). 
Starting from any vertex, there is a unique directed path of first links through a sequence of vertices. This path eventually closes when it reaches a vertex for the second time, thus creating a cycle. The subset of all the vertices connected to a given cycle forms a directed unicyclic graph. However, not all articles connect to the same cycle. The network of first links breaks down into a collection of components, each centered on a cycle towards which all first-link paths converge.

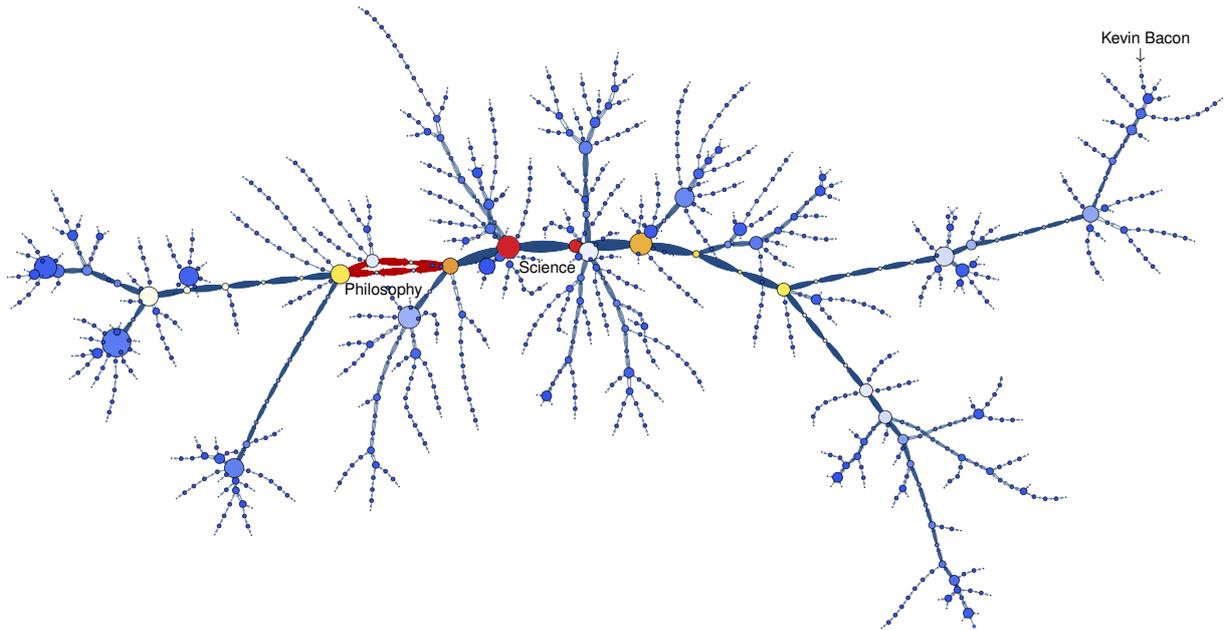
\begin{figure}[h!]
\begin{center} %\fontfamily{phv}\selectfont 
\begin{overpic}[width=\textwidth]{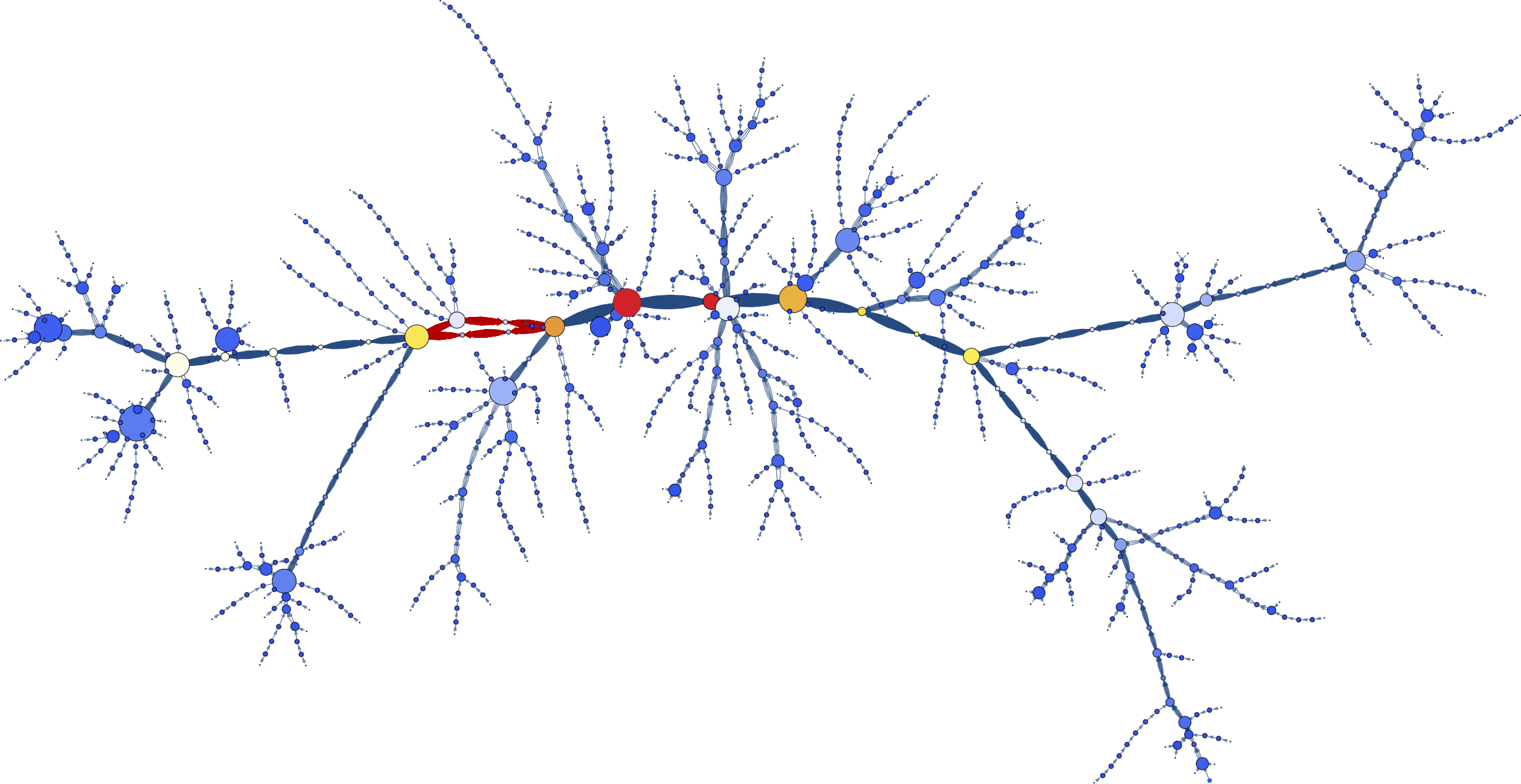}\fontfamily{phv}\selectfont 
 	\tiny	\put (42.1, 29.55) {Science}
	\put (27.75, 27.7) {Philosophy}
	\put (90, 48.5) {Kevin Bacon}
	\put (92.8, 47.2) {$\downarrow$}	
\end{overpic}
\caption{\small Network of first links on Wikipedia on August 1, 2017. Vertices represent articles and directed edges represent first links. For clarity, we restricted the network to 1,000 vertices. The core cycle was \{Philosophy, Education, Learning, Knowledge, Fact, Verificationism\} (highlighted in red). The mean distance to Philosophy was 14.3. The size of a vertex indicates its degree (number of direct neighbors), and the color temperature its betweenness centrality. Science has the highest betweenness centrality.}
\label{networkEN}
\end{center}
\end{figure}

There is in general one cycle that attracts a vast majority of vertices. This core cycle typically consists of a sequence of several vertices, whereas other cycles often only contain two vertices.
As an example, the core cycle on August 1, 2017 was made out of the concepts
\{Philosophy, Education, Learning, Knowledge, Fact, Verificationism\}
and attracted 94.3\% of the 5,395,328 articles analyzed (Figure~\ref{networkEN}). 
The largest secondary cycles were \{Canada, Provinces and territories of Canada\} (1.4\%) and \{Government, Governance\} (1.2\%) (Table~\ref{table10Cycles}).
We stress that the set of concepts in the core cycle is very unstable, as it can change drastically after a minor edit on a first link (Table~\ref{tableCycles}). Just one month earlier, on July, 1 2017, the core cycle was 
\{Knowledge, Fact, Experience\} (92.1\%) and did not include Philosophy. 
It would therefore be misleading to attach too much signification to the presence of a concept in the core cycle. Although fundamental classifying concepts are likely to be in the core cycle, peripheral concepts may enter accidentally, and some important concepts may be missing.

%%%%%%%%%%%%%%%%
\section{Hubs and scale-free distribution}

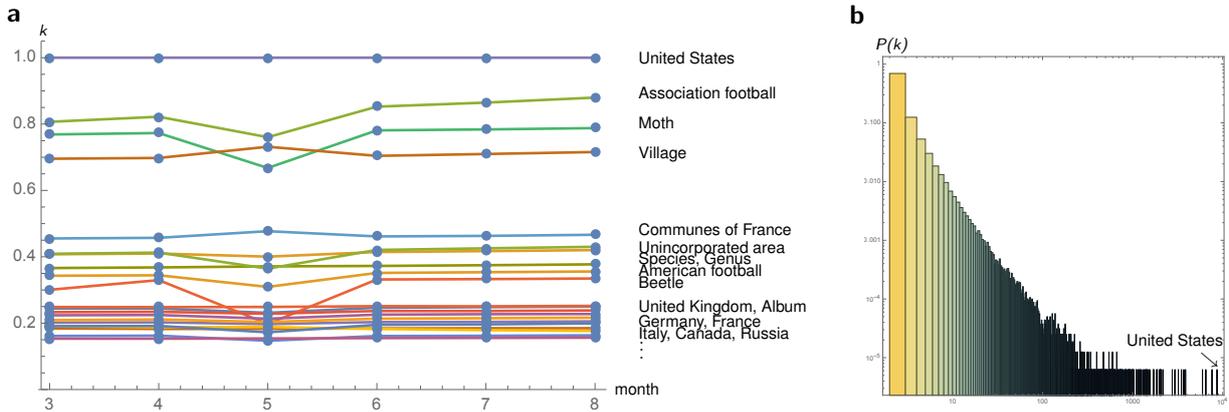
\begin{figure}[tb]
\centering
\begin{minipage}{.49\textwidth}
\centering
\begin{overpic}[width=\linewidth]{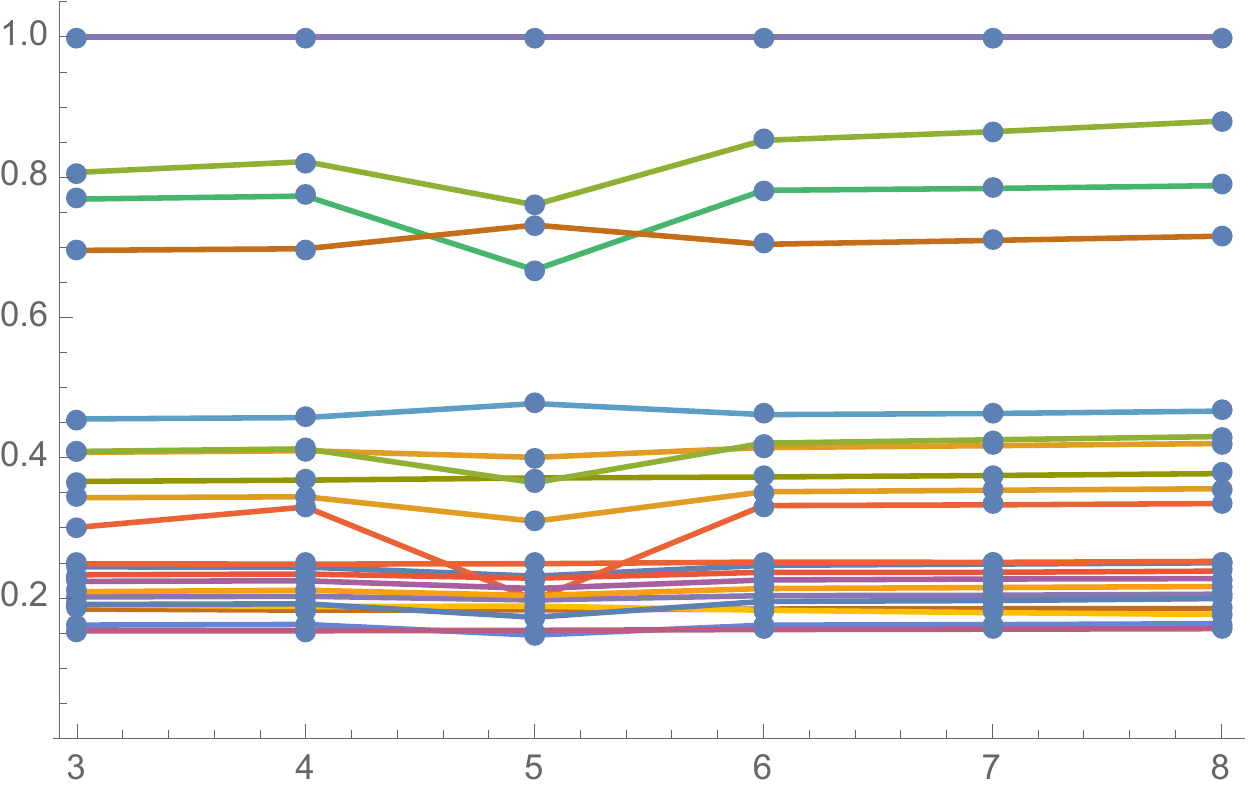}\fontfamily{phv}\selectfont 
\small \put (-1, 67) {{\bf\textsf{a}}}
\tiny	\put (4.3, 64.7) {\textsf{\textit{k}}}
	\put (105, 60) {United States}
	\put (105, 54) {Association football}
	\put (105, 49) {Moth}
	\put (105, 44) {Village}
	\put (105, 31) {Communes of France}
	\put (105, 28) {Unincorporated area}
	\put (105, 26) {Species, Genus}
	\put (105, 24) {American football}
	\put (105, 22) {Beetle}
	\put (105, 18) {United Kingdom, Album}
	\put (105, 15.5) {Germany, France}
	\put (105, 13.4) {Italy, Canada, Russia}
	\put (105, 10) {$\vdots$}
	\put (101, 4) {month}
\end{overpic}\end{minipage}\hfill
\begin{minipage}{.3\textwidth}
\centering
\begin{overpic}[width=\linewidth]{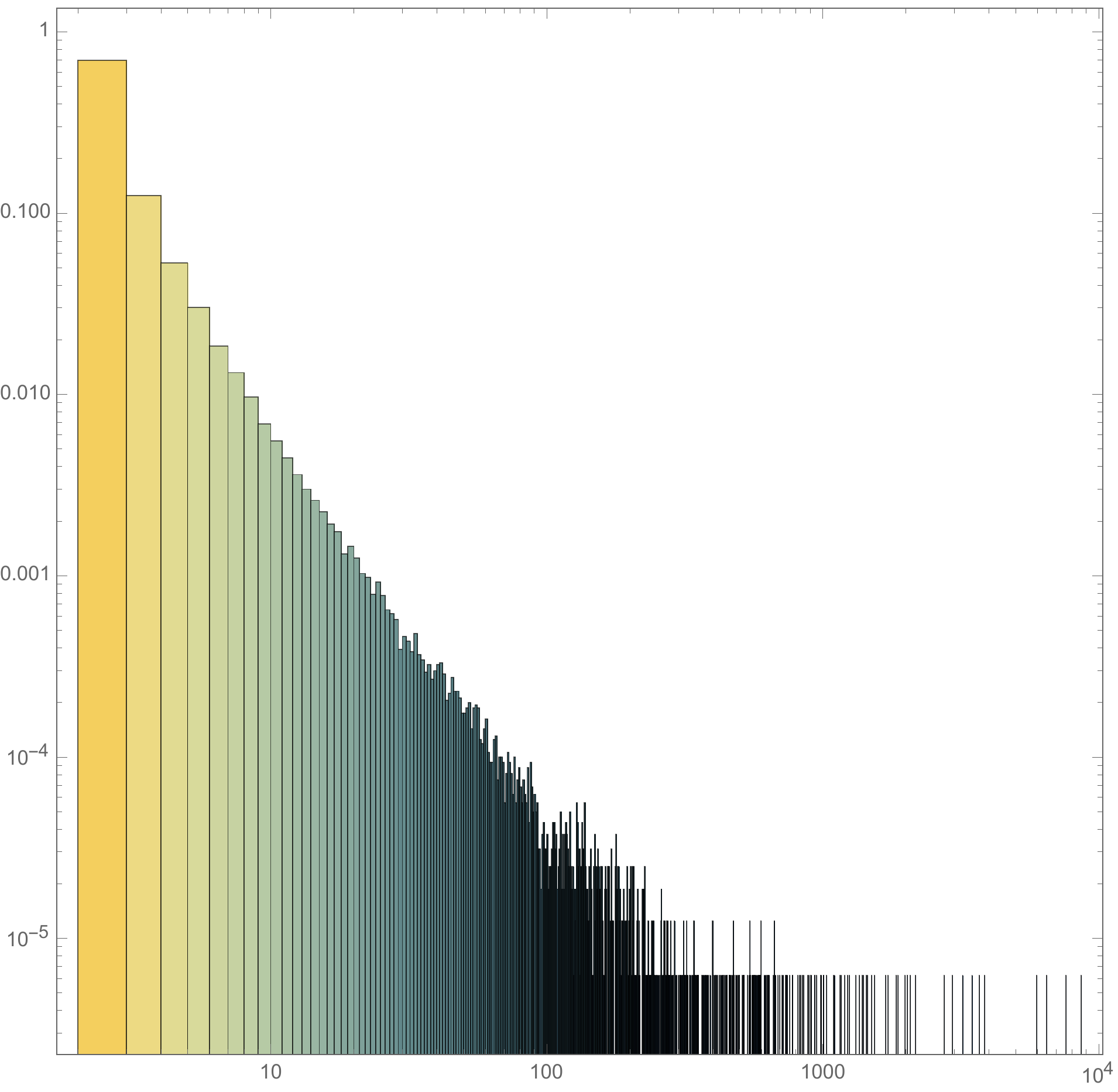}\fontfamily{phv}\selectfont 
\small \put (-4, 106) {{\bf \textsf{b}}}
\tiny	\put (3, 99) {\textsf{\textit{P(k)}}}
 	\put (72, 17) {United States}
	\put (92, 12) {$\searrow$}
\end{overpic}
\end{minipage}
\caption{\small {\bf \textsf{a}} Evolution of the highest degrees $k$ (rescaled) from March to August 2017. This measure is robust, but favors popular hubs rather than fundamental classifying concepts. {\bf \textsf{b}} Histogram of degrees (in log-log scale).  
Most vertices have very few edges, while a few vertices (hubs) have an enormous number of edges, which creates a long tail. The power-law distribution $P(k)\propto k^{-\gamma}$ (here with $\gamma \simeq 2.3$) is a universal property of scale-free networks.}
    \label{dc2017}
\end{figure}

Given the unreliability of the core cycle, we would like to find a more robust way to determine which concepts are the most fundamental.
There exist many methods in network science to measure the importance of a vertex. 
The simplest one is the degree $k$, defined as the number of incident edges
(note that by definition the out-degree of any vertex is one).
Since it would take many changes in first links to alter it significantly, this measure is very robust, as illustrated by the evolution of the highest degrees from March to August 2017 (Figure~\ref{dc2017}a). Interestingly, the degree distribution of the network of first links exhibits the power-law $P(k) \propto k^{-\gamma}$ (with $\gamma \simeq 2.3$) and long tail characteristic of scale-free networks \cite{Barabasi412}\cite{RePEc:oxp:obooks:9780199665174}\cite{1999Sci...286..509B} (Figure~\ref{dc2017}b). The degree is high for popular articles, called hubs, that appear as the first link in many articles. The predominance of the United States article is thus certainly due to the large number of American celebrities and entertainment products. Other popular categories include sports (Association football, American football), biological taxonomy (Moth, Species, Genus, Beetle), administrative divisions (Village, Commune de France, Unincorporated area), and countries (United Kingdom, Germany, France).
Despite their importance, these umbrella concepts are clearly not the most fundamental in terms of knowledge classification. 

%%%%%%%%%%%%%%%%%%%%%%%%%%%%%%
\section{Fundamental classifying concepts}

\begin{figure}[h!]
\begin{center} 
\begin{overpic}[width=0.49\textwidth]{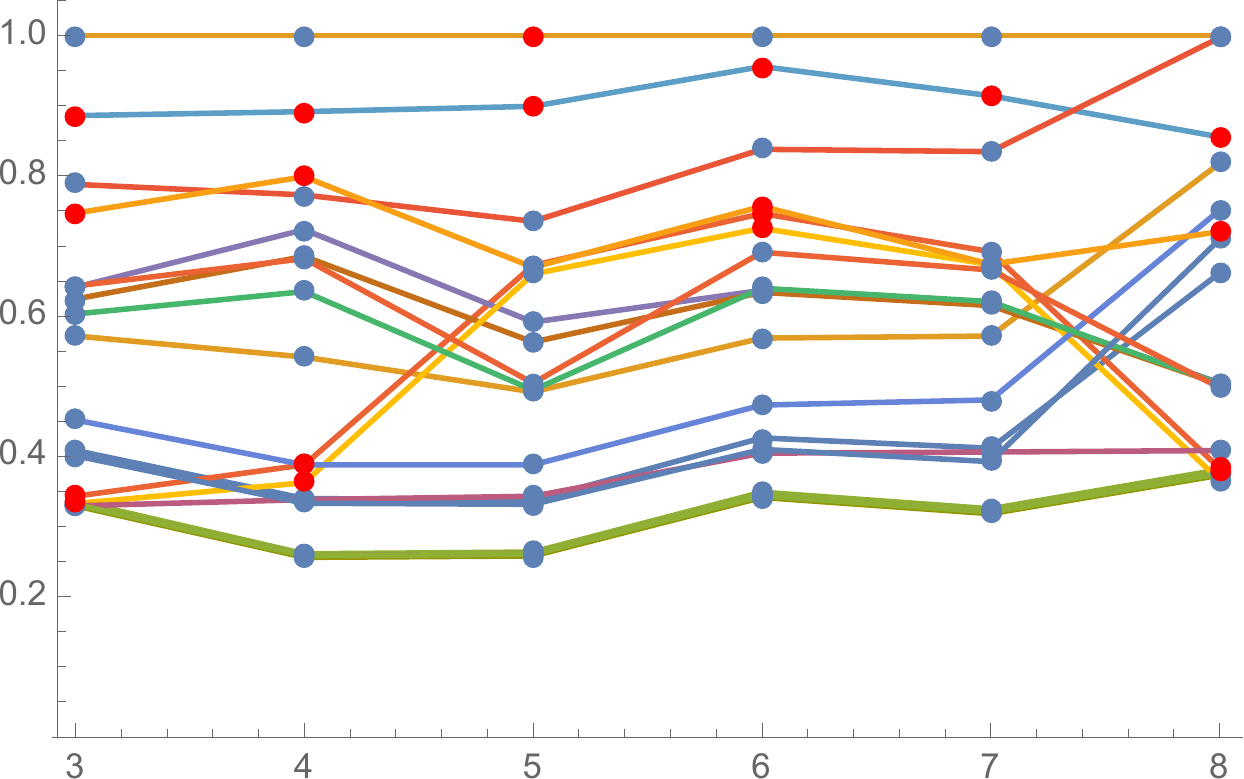}\fontfamily{phv}\selectfont 
 	\tiny	\put (3.7, 63) {\textit{C\textsubscript{B}}}
	\put (105, 60) {Science, Natural science}
	\put (105, 53) {Knowledge}
	\put (105, 50) {Biology}
	\put (105, 47) {Organism}
	\put (105, 44) {Philosophy}
	\put (105, 41.8) {Psychology}
	\put (105, 39.4) {Behavior}
	\put (105, 32.5) {State, Polity}
	\put (105, 28) {Physics}
	\put (105, 25) {Education, Learning}
	\put (105, 23) {Communication, Semiotics}
	\put (105, 19) {$\vdots$}
	\put (101, 3) {month}
\end{overpic}
\caption{\small Evolution of the highest betweenness centralities $C_B$ from March to August 2017. This robust measure is an indication of the fundamental classifying nature of a concept. Scientific concepts dominate, even when they are not among the members of the core cycle (red dots).}
\label{bcu2017}
\end{center}
\end{figure}
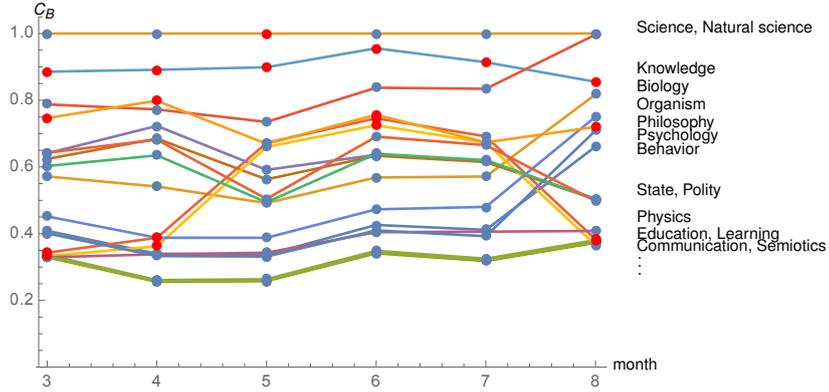

A more relevant measure of the fundamental nature of an article is its betweenness centrality $C_B(v)$ \cite{10.2307/3033543}, defined as the number of shortest paths between any pairs of vertices that run through a given vertex~$v$ (in the undirected network, so as to suppress the undesired advantage of vertices in the core cycle). Vertices with high $C_B$ are close to the core cycle or at the confluence of several important branches, which are precisely properties that we expect for fundamental concepts. 
$C_B$ is also remarkably robust (Figure~\ref{bcu2017}) and appears to be uncorrelated with the abrupt changes in the composition of the core cycle.
We indeed observe that the ranking is dominated by Science even though it was not part of the core cycle, except in May 2017. Conversely, relatively anecdotal concepts in the core cycle, such as Fact and Verificationism, do not have high $C_B$. Knowledge itself ranks high as a fundamental concept for knowledge classification.
The largest change in $C_B$ took place for the pair of articles (Education, Learning), which had a marked surge in April followed by a decline in July. It would be interesting to determine whether this is a seasonal phenomenon. 
Notwithstanding, we have found that betweenness centrality provides a robust measure of the classification importance of a concept. 

%%%%%%%%%%%%%%%%%%%%%%%%%%%%%%
\section{Cultural influence}

Our results suggest an influence of cultural history on the classification network of first links on Wikipedia. 
As mentioned above, the attractive power of the Philosophy article may be understood as a manifestation of the origin of Western civilization in the thought of ancient Greek philosophers.
In the same vein, a natural explanation for the high betweenness centrality of Science would be the growing influence of scientific thinking since the Enlightenment in Europe. 
If this is correct, we should observe significant differences for Wikipedias in languages that are not embedded in Western cultural history. We analyzed the networks of first links for several of the most well-developed Wikipedias (Figure~\ref{networkPlots}). We start by noticing that the degrees exhibit the same pattern in all languages, dominated by geographical notions, with understandable regional variations (Tables~\ref{tableDCeurop} and Figure~\ref{histograms}).
The first major difference is the composition of the core cycles (Table~\ref{tableLangCycles}). For Japanese we find  \{Human, Interpersonal relationship\} (74.7\%), and for Chinese a triplet of cycles of similar importance: \{Matter, Invariant mass, Energy, Physics\} (37.5\%), \{Solar system, Sun\} (26.9\%), and \{China, Chinese history\} (21.9\%). This is consistent with the focus on human relationship in Confucianism, one of the main influences on the East Asian cultural sphere, and with Chinese concepts such as the Qi (vital energy or material force) and the three harmonies (heaven, earth, man).
Moreover, the articles Human and Earth appear among those with the highest betweenness centrality, while scientific concepts are strikingly absent (Table~\ref{tableLanguages}, Tables~\ref{tableBCUeurop}). Note that the high betweenness centrality of some geographical hubs reflects a tendency towards popularity rather than categorization. The very presence of three giant components for Chinese illustrates the Eastern holistic approach, in contrast to the Western analytic approach.\cite{2001-17194-00120010401}\cite{doi:10.1177/0963721409359301}

\begin{table}[t]
\centering
{\fontsize{8.6}{11}\fontfamily{phv}\selectfont 
  \begin{tabular}{ c | c | c | c | c } \hline  
 \bf English  &   \bf  French &   \bf Russian &  \bf  Chinese &   \bf Japanese  \\ \hline\hline
Science  & Connaissance & State & Organism & Japan  \\   
Natural science  & Philosophie & Science  & Human &Europe   \\   
Knowledge  & Langue & Activity (process)  & Taxonomy & Eurasian continent  \\   
Biology  & Science & Organization &  Biology  & East Asia \\   
Organism  & Grec ancien & Social group  & Animal  & Eurasia  \\   
Philosophy  & Grec & Psychology& Europe   & Earth  \\   
Psychology  & Famille de langues & Activity & Bacteria  & Humanity  \\   
Behavior  & Continent & Activity (psychology)  & Gram-negative bacteria & Human \\   
Ontology  & Latin & Set (mathematics) &  Escherichia coli  & Indo-European language  \\   
Entity   & Langues italiques & Mathematics & Developmental biology   & Greek  \\ \hline 
  \end{tabular}}
  \caption{\small {\bf Concepts with highest betweenness centrality for several languages}. While Science dominates for European languages, it is outclassed for East Asian languages by concepts like Organism, Human, and Earth. 
  }\label{tableLanguages}
\end{table}

The comparison between concepts with high $C_B$ in different European languages reveals meaningful variations as well (Table~\ref{tableLanguages}). The English edition displays the values of analytic philosophy, with its emphasis on science and conceptual analysis. 
In contrast, the French edition displays the values of continental philosophy: the predominance of Philosophy over Science, and the importance of language and etymology.
It also seems significant that State dominates the Russian edition. 
We thus observe a strong correlation between concepts with high betweenness centrality and cultural heritage, in agreement with our hypothesis about the influence of culture on the structure of knowledge.

%%%%%%%%%%%%%%%%
\section{Discussion}

The discovery of a self-organizing architecture of knowledge on Wikipedia constitutes an applied philosophy demonstration of the implicit impact of cultural heritage on fundamental world views.
This phenomenon could be useful for the fully automatic construction of ontologies able to deal with constantly evolving environments\cite{BIZER2009154}\cite{SUCHANEK2008203}\cite{Ponzetto:2007:DLS:1619797.1619876}\cite{Ponzetto:2009:LTM:1661445.1661778}\cite{AllYou}\cite{Pohl2012ClassifyingTW}\cite{HOVY20132}.
An important challenge is to understand whether powerful unplanned structures also emerge in other types of data sets. This can have far-reaching implications for artificial intelligence, which would inadvertently learn deeper lessons than we would be aware of teaching.

%%%%%%%%%%%%%%%%%%%%%%%%%%%%%%%%%
%%%%%%%%%%%%%%%%%%%%%%%%%%%%%%%%%
\section{Methods}

\paragraph{Wikipedia data:}
For our analyses of Wikipedias in various languages we used mostly the `database backup dumps' on Wikimedia (\url{https://dumps.wikimedia.org}). 
Another source of Wikipedia dumps is the Internet Archive (\url{https://archive.org}).

\paragraph{First-link paths and network:}
For each Wikipedia edition, we determined the first links for all articles using a Scala program (available at~\cite{github}). Starting from any article, we constructed a path that follows first links and stops when reaching an article for the second time (we identified redirect pages with their targets). The network of first links was then assembled from such first-link paths. Note that this construction always leads to a network with mean degree $\langle k \rangle=2$. To see this, let's consider a single first-link path. Its vertices have $k=2$, expect the initial one with $k=1$ and the final one with $k=3$. Now let's add a second first-link path ending on the same cycle, which effectively amounts to gluing an open path. The degree of the vertex used for the gluing increases by one, but this is compensated by the degree $k=1$ of the added initial vertex. By recursion, we conclude that the mean degree of the full network is indeed exactly 2. 

\paragraph{Network analysis:}
We analyzed the networks of first links with Mathematica~\cite{github}. 
The degree $k$ is defined as the number of incident edges (incoming and outgoing, without multiplicity), or equivalently as the number of direct neighbors.
The betweenness centrality of a vertex $v$ is given by
$C_B(v) = \sum_{i\neq j \neq v} \frac{\sigma_{ij}(v)}{\sigma_{ij}}$,
where $\sigma_{ij}$ is the total number of shortest paths from vertex $i$ to vertex $j$, and $\sigma_{ij}(v)$ is the number of those that pass through $v$. 
Although networks of first links are directed, we used betweenness centrality for their undirected versions as a measure of the fundamental classifying nature of vertices.

\paragraph{Network visualization:}
When plotting a network of first links, we restricted for clarity to a small subset of the giant component, constructed by superimposing first-link paths until we obtained 1,000 vertices. The multiplicity of an edge gives a visual indication of the number of paths that run through it. This reveals the circulatory structure of the network, with the core cycle at its heart.

%%%%%%%%%%%%%%%%%%%%%%%%%%%%%%%%%
\section*{Acknowledgements}

I would like to thank Maurice Zomorrodi for inspiration and Michael Gottschalk for programing assistance. 
I also thank Emmanuel Abbe, Atif Ansar, Jean-Paul de Vooght, and Charlotte F.~Werbe for comments on the manuscript.
This research was supported by the Swiss National Science Foundation (project P300P2-158440).

%%%%%%%%%%%%%%%%%%%%%%%%%%%%%%%%%
%%%%%%%%%%%%%%%%%%%%%%%%%%%%%%%%%

\clearpage
\appendix

%%%%%%%%%%%%%%%%%%%%%%%%%%%%%%%%
%%%%%%%%%%%%%%%%%%%%%%%%%%%%%%%%
\section{Figures}

\begin{figure}[h!]
\centering
\begin{minipage}{.49\textwidth}
\centering
\begin{overpic}[width=\linewidth]{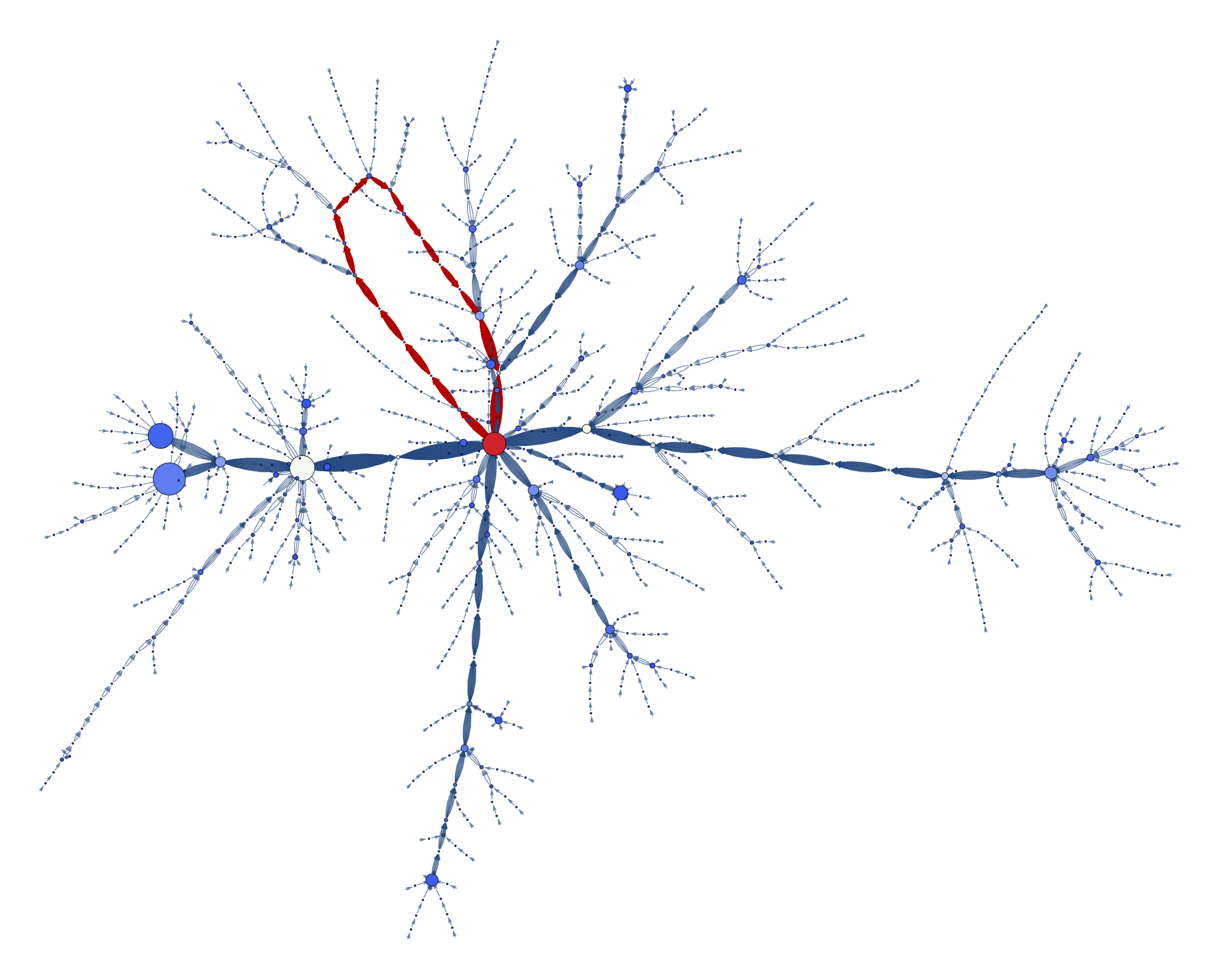}\fontfamily{phv}\selectfont 
 	\tiny	\put (5, 80) {{\bf\textsf{a}} German}
	\put (25, 44.9) {Wissenschaft}
	%\put (28.3, 27.6) {$\searrow$}
\end{overpic}\end{minipage}\hfill
\begin{minipage}{.49\textwidth}
\centering
\begin{overpic}[width=\linewidth]{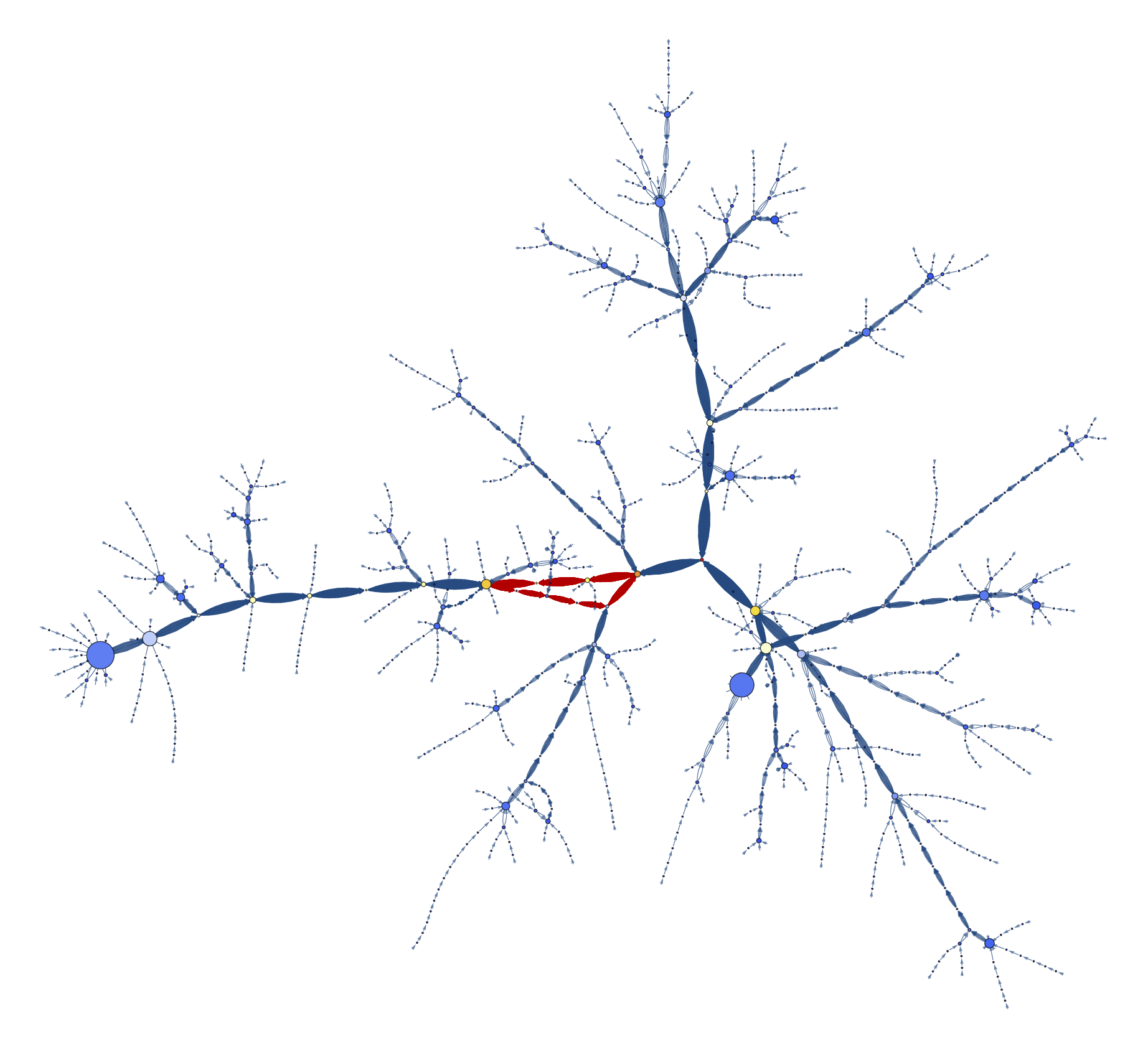}\fontfamily{phv}\selectfont 
 	\tiny	\put (5, 85) {{\bf \textsf{b}} French}
	\put (62.2, 42.6) {Connaissance}
	%\put (28.3, 27.6) {$\searrow$}
\end{overpic}
\end{minipage}\hfill
\begin{minipage}{.37\textwidth}
\centering
\begin{overpic}[width=\linewidth]{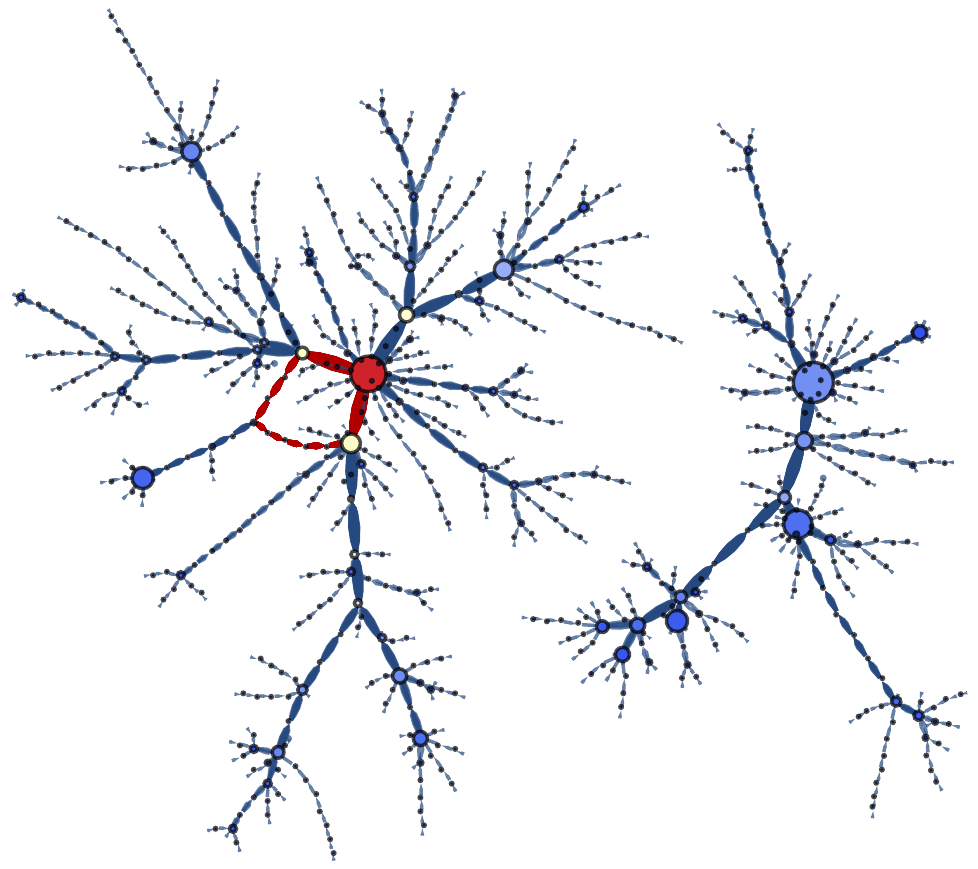}\fontfamily{phv}\selectfont 
 	\tiny	\put (5, 90) {{\bf \textsf{c}} Italian}
	\put (42.3, 51.9) {Scienza}
	%\put (28.3, 27.6) {$\searrow$}
\end{overpic}
\end{minipage}
\begin{minipage}{.61\textwidth}
\centering
\begin{overpic}[width=\linewidth]{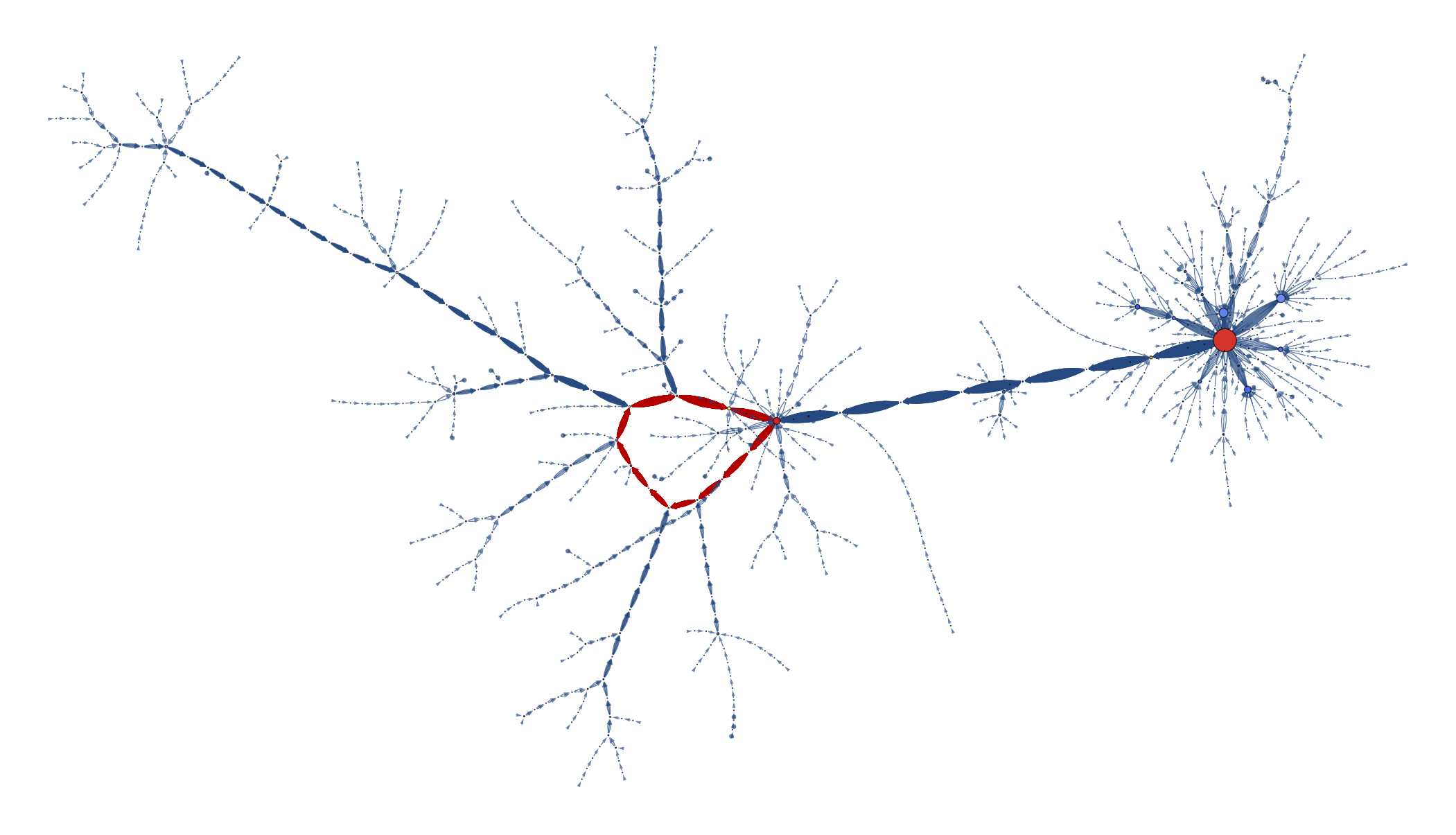}\fontfamily{phv}\selectfont 
 	\tiny	\put (5, 56) {{\bf \textsf{d}} Russian}
	\put (88.5, 34.2) {State}
	%\put (28.3, 27.6) {$\searrow$}
\end{overpic}\end{minipage}\hfill
\begin{minipage}{.49\textwidth}
\centering
\begin{overpic}[width=\linewidth]{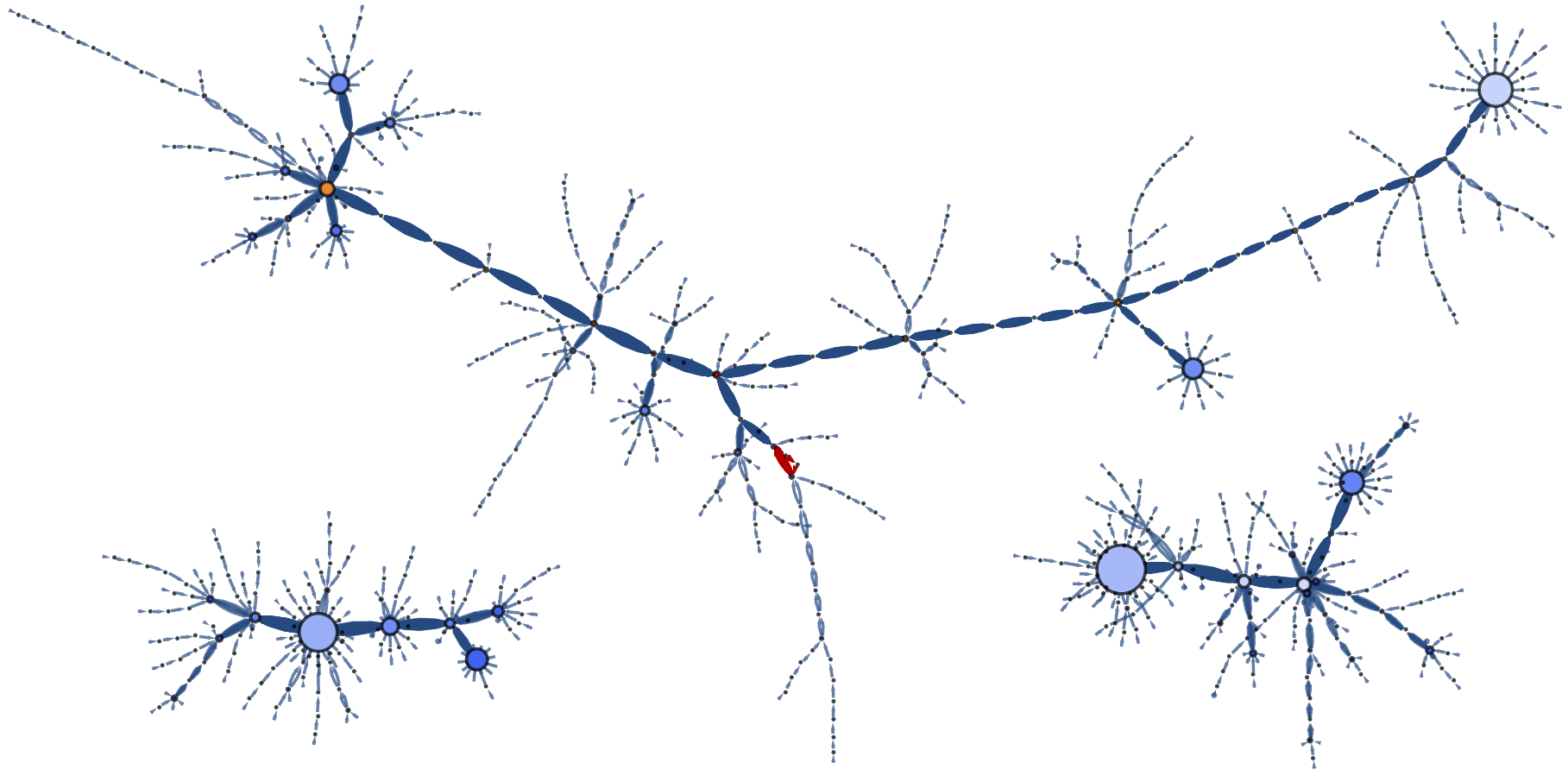}\fontfamily{phv}\selectfont 
 	\tiny	\put (5, 58) {{\bf \textsf{e}} Chinese}
	\put (44, 29.8) {Organism}
	\put (44.9, 27.2) {$\downarrow$}
\end{overpic}\end{minipage}\hfill
\begin{minipage}{.49\textwidth}
\centering
\begin{overpic}[width=\linewidth]{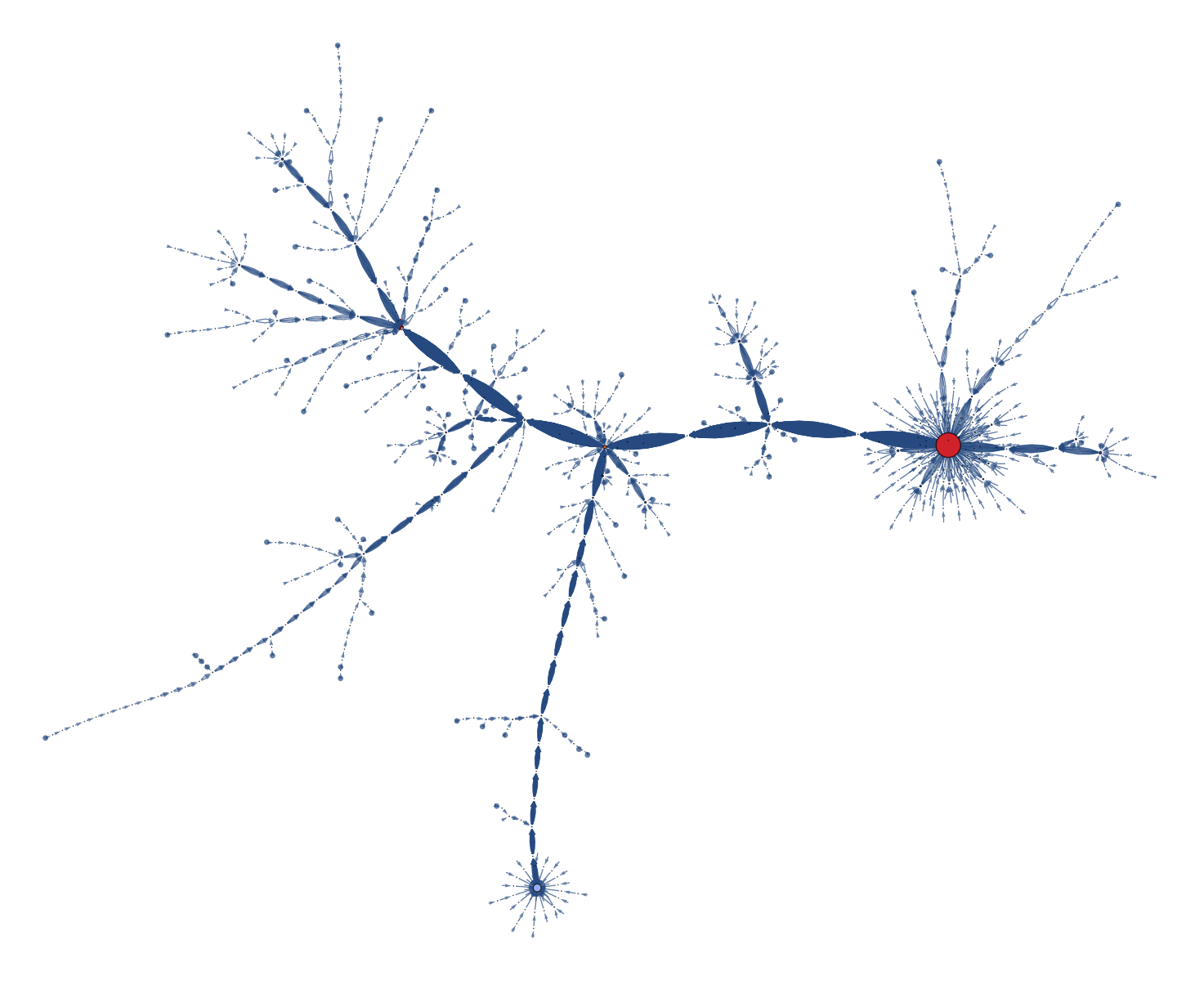}\fontfamily{phv}\selectfont 
 	\tiny	\put (5, 75) {{\bf \textsf{f}} Japanese}
	\put (76, 36) {Japan}
	%\put (28.3, 27.6) {$\searrow$}
\end{overpic}\end{minipage}
\caption{\small Networks of first links on Wikipedia around the core cycle (highlighted in red) for several languages. We show all components with more than 20\% of all vertices. For European languages, the networks are intensely ramified and centered on vertices with highest centrality (indicated). In contrast, for East Asian languages, hubs dominate and the ramification is uneven. The Russian network present hybrid characteristics.
}
    \label{networkPlots}
\end{figure}

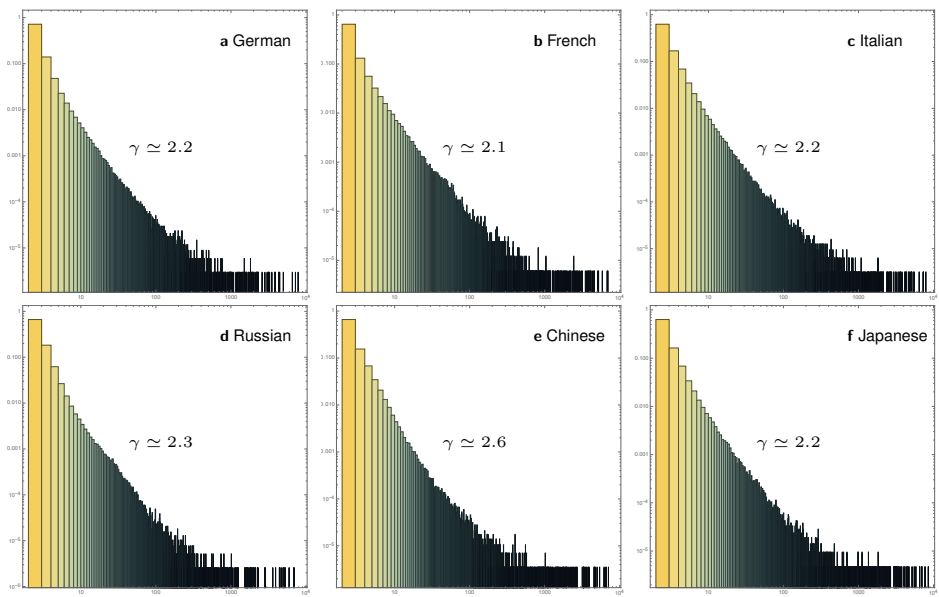
\begin{figure}[h!]
\begin{center} 
\begin{overpic}[width=0.25\textwidth]{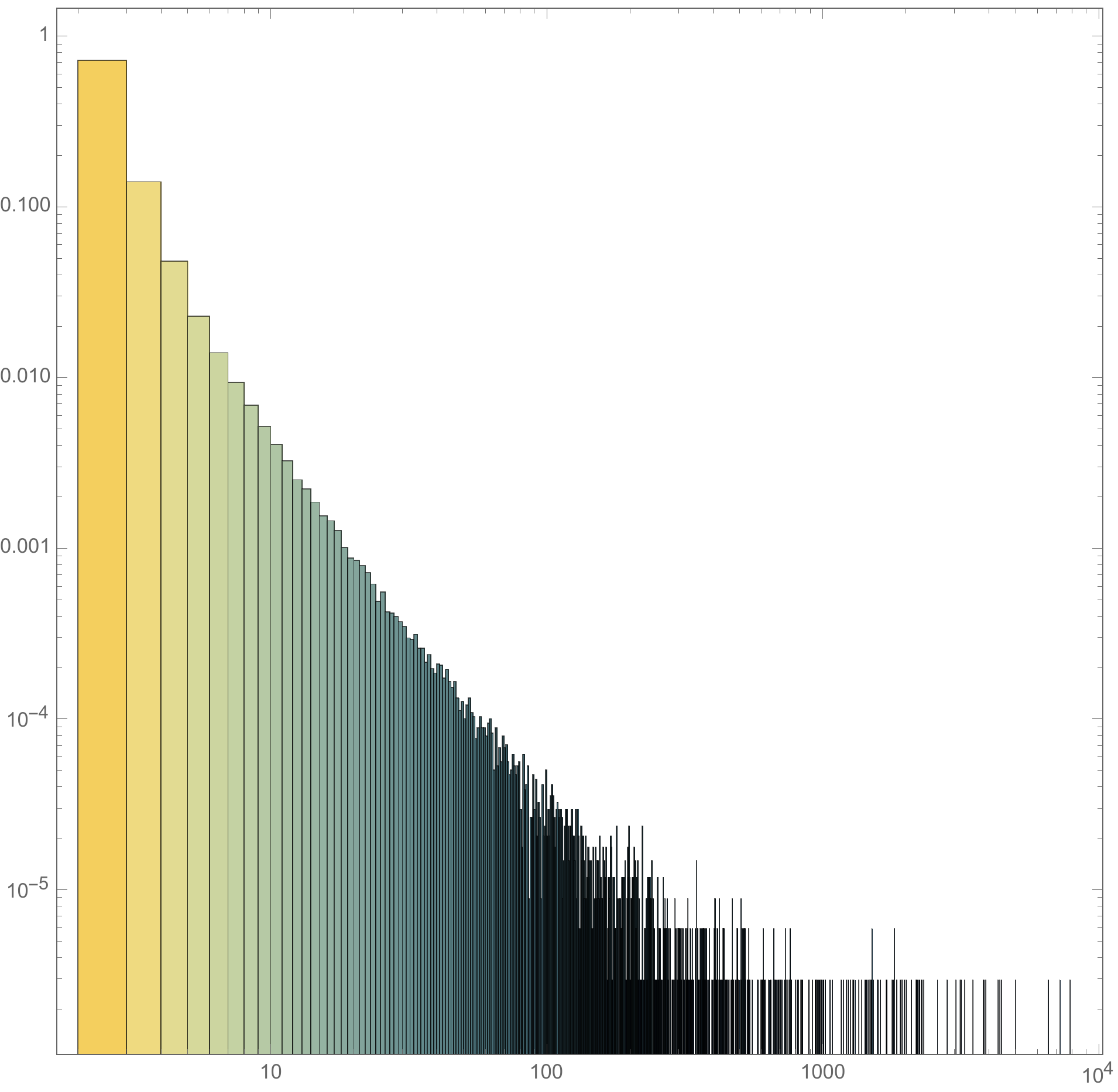}\fontfamily{phv}\selectfont 
 	\tiny	\put (70, 85) {{\bf\textsf{a}} German}
	\put (40, 50) {$\gamma \simeq 2.2$}
\end{overpic}
\begin{overpic}[width=0.25\textwidth]{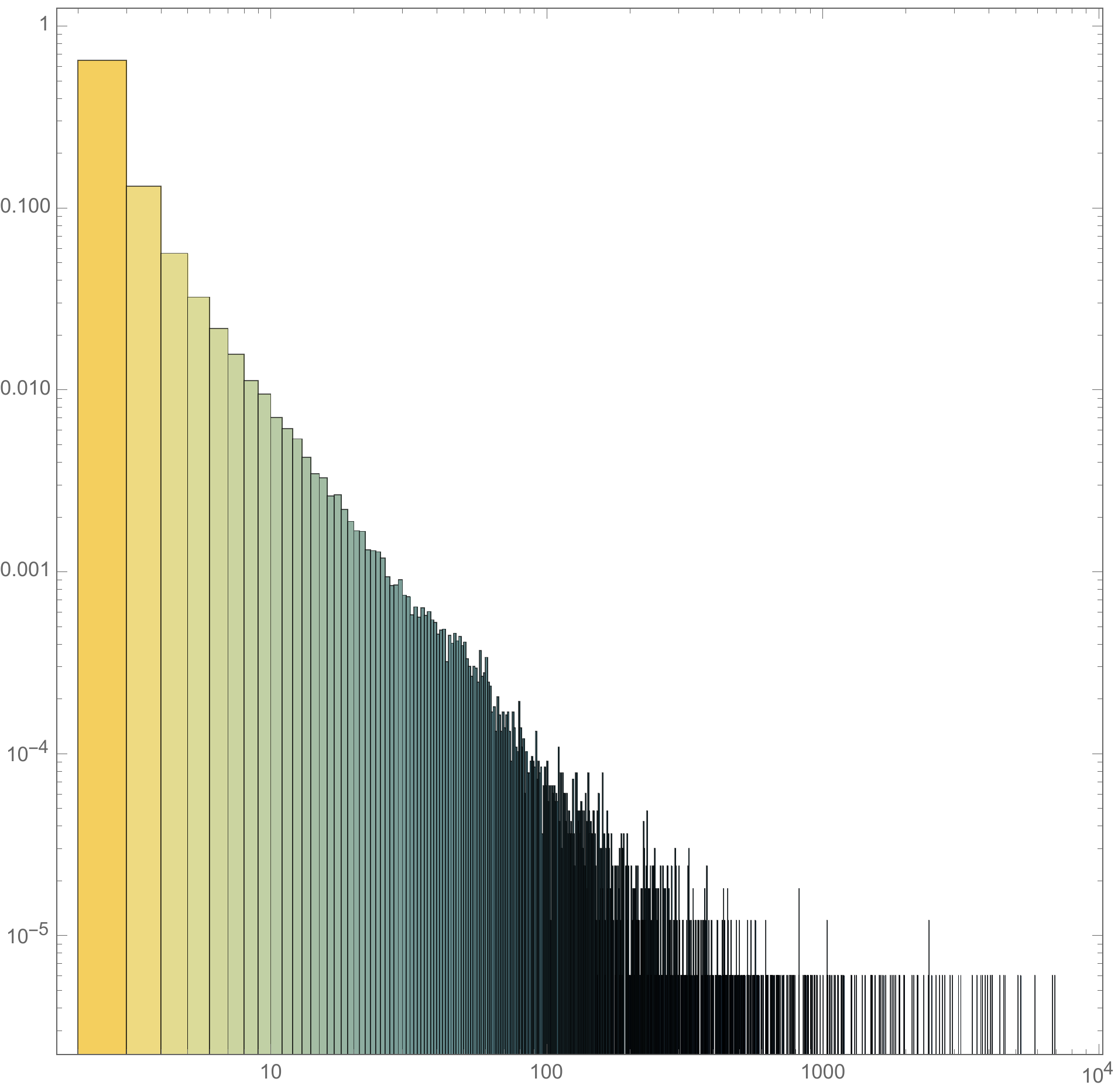}\fontfamily{phv}\selectfont 
 	\tiny	\put (70, 85) {{\bf\textsf{b}} French}
	\put (40, 50) {$\gamma \simeq 2.1$}
\end{overpic}
\begin{overpic}[width=0.25\textwidth]{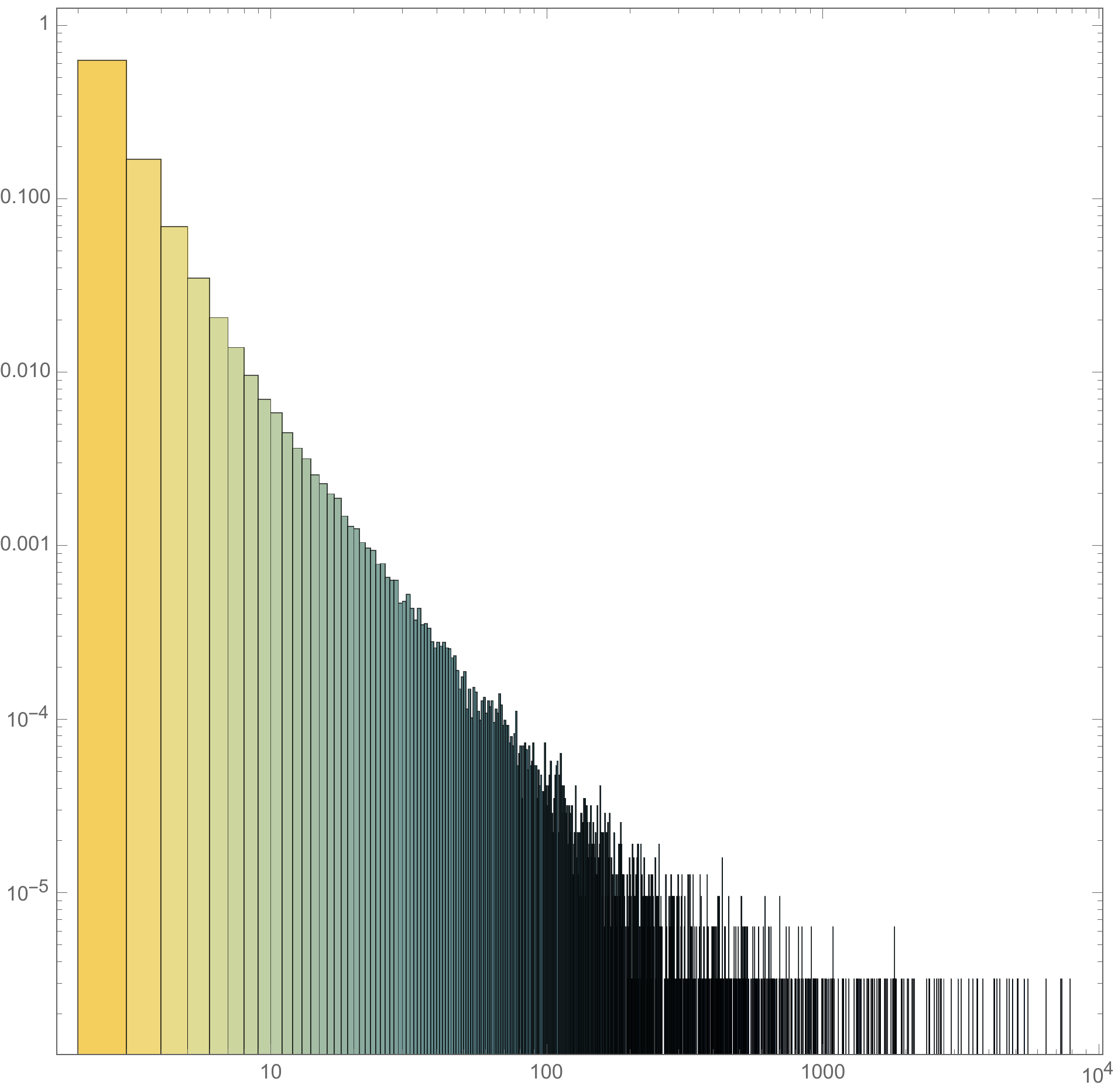}\fontfamily{phv}\selectfont 
 	\tiny	\put (70, 85) {{\bf\textsf{c}} Italian}
	\put (40, 50) {$\gamma \simeq 2.2$}
\end{overpic}
\begin{overpic}[width=0.25\textwidth]{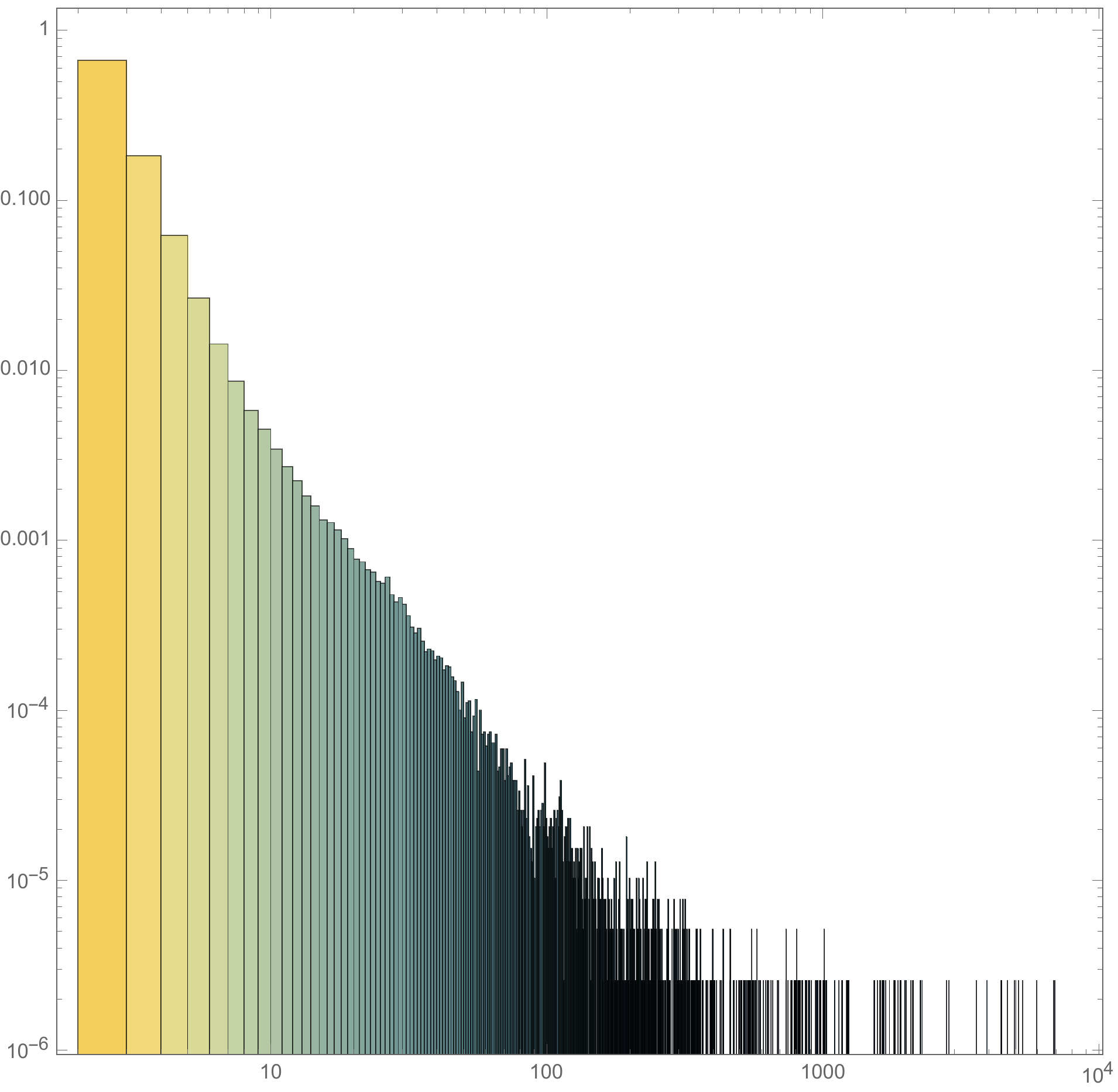}\fontfamily{phv}\selectfont 
 	\tiny	\put (70, 85) {{\bf\textsf{d}} Russian}
	\put (40, 50) {$\gamma \simeq 2.3$}
\end{overpic}
\begin{overpic}[width=0.25\textwidth]{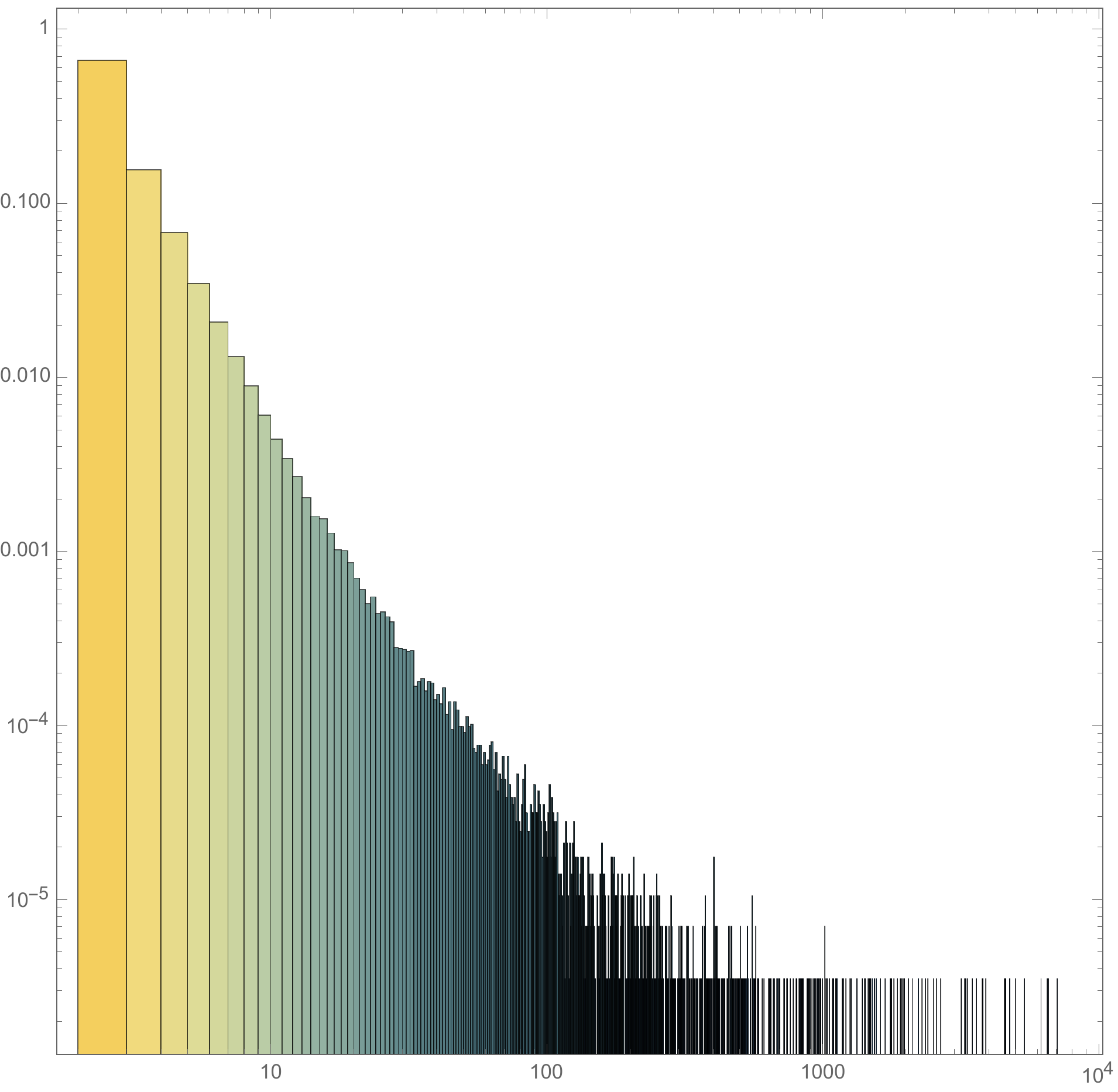}\fontfamily{phv}\selectfont 
 	\tiny	\put (70, 85) {{\bf\textsf{e}} Chinese}
	\put (40, 50) {$\gamma \simeq 2.6$}
\end{overpic}
\begin{overpic}[width=0.25\textwidth]{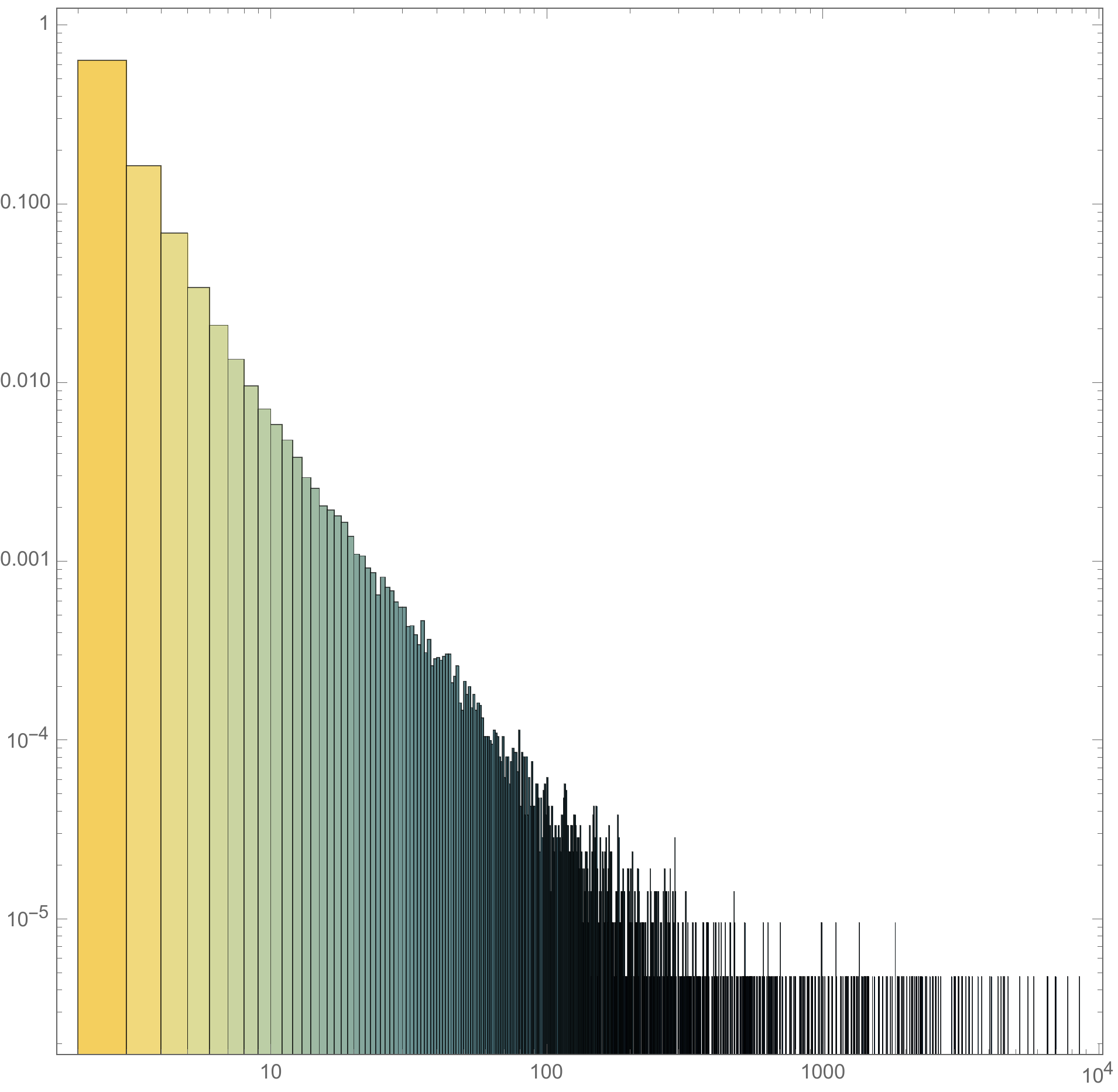}\fontfamily{phv}\selectfont 
 	\tiny	\put (70, 85) {{\bf\textsf{f}} Japanese}
	\put (40, 50) {$\gamma \simeq 2.2$}
\end{overpic}
\caption{\small Histogram of degrees (in log-log scale) in several languages. Most vertices have very few edges, while a few vertices (hubs) have an enormous number of edges, which creates a long tail. The power-law distribution $P(k)\propto k^{-\gamma}$ is a universal property of scale-free networks, here with $2.1\lesssim \gamma \lesssim 2.6$.  }
\label{histograms}
\end{center}
\end{figure}

\clearpage
%%%%%%%%%%%%%%%%%%%%%%%%%%%%%%%%
%%%%%%%%%%%%%%%%%%%%%%%%%%%%%%%%
\section{Tables}

\begin{table}[h]
\centering
{\fontsize{7}{9}\fontfamily{phv}\selectfont 
\begin{tabular}{ p{7.3cm} | c }\hline
Cycle & Size   \\ \hline 
 Knowledge, Fact, Verificationism, Philosophy, Education, Learning   &  94.28\%  \\ 
 Canada, Provinces and territories of Canada   &  1.39\%  \\ 
 Government, Governance   &  1.20\%  \\ 
 Basketball, Basketball court    &  0.36\%  \\ 
Telecommunication, Transmission (telecommunications)    &  0.23\%  \\ 
Southern United States, Dixie    &  0.18\%  \\ 
 January 1, New Year's Day    &  0.11\%  \\ 
  Finance, Investment, Rate of return   &  0.11\%  \\ 
Port, Harbor    &  0.10\%  \\ 
 Monotheism, God   &  0.08\%  \\ \hline
\end{tabular}}
\caption{\small {\bf The ten largest cycles for the English Wikipedia on August 1, 2017}. The core cycle in the giant component involves fundamental concepts, whereas the secondary cycles mostly consist of pairs of incidentally linked concepts.}
\label{table10Cycles}
\end{table}

\begin{table}[h]
\centering
{\fontsize{7}{9}\fontfamily{phv}\selectfont 
\begin{tabular}{ c | p{12cm} | c  }\hline
Date & Core cycle & Size   \\ \hline 
%2010 &  Science, System, Entity, Existence, Sense, Perception, Philosophy, Reason, Mind, Intelligence, Intelligence (information gathering), Information, Conveyed concept, Set phrase, Phrase, Grammar, Linguistics  &  72\%   \\ 
20170301 &   Knowledge, Awareness, Quality (philosophy), Philosophy, Education, Learning  &  96.1\%   \\ 
20170401 &  Knowledge, Awareness, Quality (philosophy), Philosophy, Education, Learning   &    95.7\%   \\ 
20170501 &  Science, Knowledge, Awareness, Perception, Information, Question, Referring expression, Linguistics   &  91.4\%  \\ 
20170601 &  Knowledge, Fact, Axiom, Premise, Logic, Argument, Philosophy, Education, Learning   &  93.1\%  \\ 
20170701 &  Knowledge, Fact, Experience   &  92.1\%  \\ 
20170801 &  Knowledge, Fact, Verificationism, Philosophy, Education, Learning   &  94.3\%  \\ \hline
\end{tabular}}
\caption{\small {\bf Core cycle for the English Wikipedia from March to August 2017}. Knowledge is the only article that entered in all the core cycles, while Philosophy was absent in May and July. Science was only present in May.}
\label{tableCycles}
\end{table}

%Table of core cycles for Wikipedia in several languages (on 20170701):
%%
%\begin{center}
%\begin{tabular}{| c | p{10cm} | p{1cm} | c |}\hline
%Language & Core cycle & Size  & Total \\ \hline \hline
%EN-20170701 &  Knowledge, Fact, Experience   &  92.1\% & 5,373,975 \\ \hline
%DE-20170701 &   Wissenschaft, Wissen, Gewissheit, Sicherheit, Risiko, Ereignis, Beobachtung, Ziel, Zukunft, Zeit, Physikalische Größe, Quantifizierung, Eigenschaft, Person, Sozialphilosophie, Gesellschaft (Soziologie), Soziologie  &  84.7\% & 2,619,965  \\ \hline
%FR-20170701 &  Philosophie, Grec ancien, Grec, Langue, Système, Ensemble, Axiome, Proposition (philosophie)    &  93.4\% & 1,867,968  \\ \hline
%IT-20170701 &  Biologia, Scienza, Conoscenza, Consapevolezza, Psicologia, Psiche, Cervello, Sistema nervoso centrale, Sistema nervoso, Tessuto (biologia) ---  Diritto, Stato, Persona giuridica &  53.9\% 34.0\% & 1,547,468  \\ \hline
%RU-20170701 &  Science, Objectivity, Object (philosophy), Category (philosophy), Generalization of concepts, Logical operation, Logic, Philosophy, Cognition, Plurality, Mathematics   &  92.0\% & 2,176,658 \\ \hline
%JA-20170701 &     Human being, Interpersonal relationship --- Number, Ordered set, Mathematics, Quantity   &   78.9\% 10.5\% &  1,261,704  \\ \hline
%ZH-20170701 & Solar system, Sun ---  Matter, Invariant mass, Energy, Physics  &  36\%       30\% & 1,307,410  \\ \hline
%\end{tabular}
%\end{center}
%%

%
\begin{table}[h]
\centering
{\fontsize{7}{9}\fontfamily{phv}\selectfont 
\begin{tabular}{ l | p{9.1cm} | c | c | c }\hline
Language & Core cycles & $N_\text{tot}$ & $\langle \ell \rangle$ & $\gamma$ \\ \hline \hline
English &  Knowledge, Fact, Verificationism, Philosophy, Education, Learning (94.3\%) & 5,395,328   & 18.5 & 2.3  \\ \hline
German &   Wissenschaft, Wissen, Gewissheit, Sicherheit, Risiko, Ereignis, Beobachtung, Ziel, Zukunft, Zeit, Physikalische Größe, Quantifizierung, Eigenschaft, Person, Sozialphilosophie, Gesellschaft (Soziologie), Soziologie (84.7\%) & 2,619,965  & 24.3 & 2.2  \\
 &   Philosophie, Welt, Totalit\"at (12.1\%) &    & &  \\ \hline
French &   Philosophie, Grec ancien, Grec, Langue, Système, Ensemble, Axiome, Proposition (philosophie) (89.7\%) & 1,878,265  & 17.8  & 2.1  \\ \hline
Italian &   Biologia, Scienza, Conoscenza, Consapevolezza, Psicologia, Psiche, Cervello, Sistema nervoso centrale, Sistema nervoso, Tessuto (biologia)  (53.6\%) & 1,556,670 &  14.6  & 2.2  \\  
 &     Diritto, Stato, Persona giuridica (34.2\%) &  &    &   \\ \hline
Russian &   Science, Objectivity, Object (philosophy), Category (philosophy), Generalization of concepts, Logical operation, Logic, Philosophy, Cognition, Plurality, Mathematics  (92.3\%)& 2,177,505 &  22.5  & 2.3  \\ \hline
Chinese &  Matter, Invariant mass, Energy, Physics (37.5\%)   & 1,311,009 &   11.9 & 2.6  \\
  &  Solar system, Sun (26.9\%) &   &   &   \\ 
  &  China, Chinese history (21.9\%) &   &   &   \\ \hline
  Japanese &  Human, Interpersonal relationship (74.7\%)  &  1,267,451 & 11.9  & 2.2   \\
 &   Number, Ordered set, Mathematics, Quantity (10.5\%) &   &   &    \\ \hline
\end{tabular}}
\caption{\small {\bf Core cycles for Wikipedias in several languages on August 1, 2017}.  We also indicated secondary cycles of non-negligible size. $N_\text{tot}$ is the total number of articles analyzed, $\langle \ell \rangle$ the average length of first-link paths, and $\gamma$ the scale-free exponent. }
\label{tableLangCycles}
\end{table}

\begin{table}[h!]
\centering
{\fontsize{7}{9}\fontfamily{phv}\selectfont 
  \begin{tabular}{cc|cc|cc|cc} \hline  
 English &  \textit{k}& German & \textit{k} &   French &  \textit{k} & Italian & \textit{k}  \\ \hline
\text{United States} & 1.0 & \text{Vereinigte Staaten} & 1.0 & \text{Esp{\` e}ce} & 1.0 & \text{Comuni della Francia} & 1.0 \\
 \text{Association football} & 0.88 & \text{Deutschland} & 0.67 & \text{Commune (France)} & 0.80 & \text{Film} & 0.65 \\
 \text{Moth} & 0.79 & \text{Frankreich} & 0.62 & \text{Football} & 0.43 & \text{Tennis} & 0.61 \\
 \text{Village} & 0.72 & \text{{\" O}sterreich} & 0.33 & \text{Ast{\' e}ro{\" \i}de} & 0.35 & \text{Asteroide} & 0.57 \\
 \text{Communes of France} & 0.47 & \text{Italien} & 0.24 & \text{Genre (biologie)} & 0.35 & \text{Album discografico} & 0.52 \\
 \text{Unincorporated area} & 0.43 & \text{Schweiz} & 0.22 & \text{Acteur} & 0.22 & \text{Stati Uniti d'America} & 0.51 \\
 \text{Species} & 0.42 & \text{Vereinigtes K{\" o}nigreich} & 0.20 & \text{{\' E}tats-Unis} & 0.22 & \text{Cortometraggio} & 0.45 \\
 \text{Genus} & 0.38 & \text{Russland} & 0.15 & \text{Village} & 0.19 & \text{Comuni della Spagna} & 0.22 \\
 \text{American football} & 0.36 & \text{Art (Biologie)} & 0.14 & \text{France} & 0.17 & \text{Serie televisiva} & 0.20 \\
 \text{Beetle} & 0.33 & \text{R{\" o}misch-katholische Kirche} & 0.14 & \text{Paris} & 0.16 & \text{Comune (Italia)} & 0.20 \\
 \text{United Kingdom} & 0.25 & \text{Kanada} & 0.13 & \text{S{\' e}rie t{\' e}l{\' e}vis{\' e}e} & 0.15 & \text{Comuni della Germania} & 0.18 \\
 \text{Album} & 0.25 & \text{Schauspieler} & 0.12 & \text{Commune (Italie)} & 0.15 & \text{Germania} & 0.15 \\
 \text{Germany} & 0.24 & \text{Polen} & 0.12 & \text{Personnalit{\' e} politique} & 0.14 & \text{Chiesa cattolica} & 0.15 \\
 \text{France} & 0.23 & \text{Schweden} & 0.12 & \text{Cin{\' e}ma am{\' e}ricain} & 0.13 & \text{Stazione ferroviaria} & 0.14 \\
 \text{Italy} & 0.22 & \text{Japan} & 0.11 & \text{Bosnie-Herz{\' e}govine} & 0.13 & \text{Gruppo musicale} & 0.14 \\
 \text{Canada} & 0.21 & \text{Spanien} & 0.10 & \text{Polonais (peuple)} & 0.13 & \text{Singolo (musica)} & 0.14 \\
 \text{Russia} & 0.20 & \text{Denkmalschutz} & 0.098 & \text{Pologne} & 0.13 & \text{Codice aeroportuale IATA} & 0.13 \\
 \text{Canadians} & 0.18 & \text{Fu{\ss}ball} & 0.098 & \text{Cin{\' e}ma fran{\c c}ais} & 0.12 & \text{India} & 0.13 \\
 \text{England} & 0.18 & \text{Niederlande} & 0.097 & \text{Cyclisme} & 0.12 & \text{Comuni della Repubblica Ceca} & 0.13 \\
 \text{Basketball} & 0.17 & \text{Tschechien} & 0.094 & \text{Jeu vid{\' e}o} & 0.12 & \text{Cantoni della Francia} & 0.12 \\
% \text{Australian rules football} & 0.17 & \text{Australien} & 0.087 & \text{Athl{\' e}tisme} & 0.12 & \text{Squadra di calcio} & 0.12 \\
% \text{India} & 0.16 & \text{Deutsche} & 0.081 & \text{Gare ferroviaire} & 0.11 & \text{Film muto} & 0.11 \\
% \text{Japan} & 0.16 & \text{Bischof} & 0.072 & \text{Commune (Espagne)} & 0.10 & \text{Cratere meteoritico} & 0.11  \\ 
 \hline 
  \end{tabular}
  
  \vspace{.5cm}

\begin{tabular}{cc|cc|cc} \hline  
 Russian & \textit{k} & Chinese & \textit{k} &   Japanese & \textit{k}    \\ \hline
 \text{Russia}  & 1.0 & \text{France} & 1.0 & \text{Japan} & 1.0 \\
 \text{United States of America} & 0.64 & \text{Township} & 0.96 & \text{United States} & 0.47 \\
 \text{USSR} & 0.56 & \text{United States} & 0.71 & \text{Tokyo} & 0.19 \\
 \text{Germany} & 0.51 & \text{Japan} & 0.67 & \text{Italy} & 0.15 \\
 \text{Village} & 0.36 & \text{Ukraine} & 0.66 & \text{China} & 0.11 \\
 \text{Italy} & 0.36 & \text{Finfish} & 0.61 & \text{Germany} & 0.11 \\
 \text{Commune of France} & 0.31 & \text{China} & 0.61 & \text{United Kingdom} & 0.11 \\
 \text{Football} & 0.27 & \text{Germany} & 0.41 & \text{France} & 0.11 \\
 \text{Hamlet} & 0.27 & \text{Hong Kong} & 0.40 & \text{Aichi prefecture} & 0.11 \\
 \text{France} & 0.26 & \text{People 's Republic of China} & 0.38 & \text{Hokkaido} & 0.10 \\
 \text{Ukraine} & 0.26 & \text{Antarctica} & 0.34 & \text{Osaka prefecture} & 0.095 \\
 \text{Municipal division} & 0.24 & \text{Italy} & 0.31 & \text{Edo period} & 0.086 \\
 \text{Municipality} & 0.23 & \text{Sun} & 0.30 & \text{Hyogo prefecture} & 0.078 \\
 \text{Brazil} & 0.23 & \text{Spain} & 0.28 & \text{Kanagawa prefecture} & 0.078 \\
 \text{United Kingdom} & 0.20 & \text{Taiwan} & 0.26 & \text{Saitama ken} & 0.073 \\
 \text{Bulgaria} & 0.19 & \text{Russia} & 0.26 & \text{South Korea} & 0.073 \\
 \text{Portugal} & 0.15 & \text{Brazil} & 0.19 & \text{Chiba prefecture} & 0.065 \\
 \text{City} & 0.15 & \text{India} & 0.18 & \text{China} & 0.062 \\
 \text{Austria} & 0.12 & \text{United Kingdom} & 0.18 & \text{Fukuoka prefecture} & 0.058 \\
 \text{Species} & 0.12 & \text{Taiwan} & 0.17 & \text{Gifu prefecture} & 0.052 \\
% \text{Great Patriotic War} & 0.12 & \text{Galactic coordinate system} & 0.15 & \text{Shizuoka prefecture} & 0.051 \\
% \text{River} & 0.11 & \text{Ming dynasty} & 0.14 & \text{Niigata prefecture} & 0.050 \\
% \text{Japan} & 0.11 & \text{Guangdong Province} & 0.13 & \text{Russia} & 0.049    \\ 
 \hline 
  \end{tabular}
  }
  \caption{\small {\bf Concepts with highest degree for several European and East Asian languages}. Geographical notions dominate broadly, with understandable regional variations.
  }\label{tableDCeurop}
\end{table}

\begin{table}[h!]
\centering
{\fontsize{7}{9}\fontfamily{phv}\selectfont 
  \begin{tabular}{cc|cc|cc|cc} \hline  
 English & \textit{C\textsubscript{B}} & German & \textit{C\textsubscript{B}} &   French &  \textit{C\textsubscript{B}} & Italian &  \textit{C\textsubscript{B}} \\ \hline
 \text{Science} & 1.0 & \text{Wissenschaft} & 1.0 & \text{Connaissance} & 1.0 & \text{Scienza} & 1.0 \\
 \text{Natural science} & 1.0 & \text{Staat} & 0.60 & \text{Philosophie} & 0.87 & \text{Linguistica} & 0.56 \\
 \text{Knowledge} & 0.86 & \text{Sozialwissenschaften} & 0.53 & \text{Langue} & 0.78 & \text{Conoscenza} & 0.55 \\
 \text{Biology} & 0.82 & \text{Naturwissenschaft} & 0.40 & \text{Science} & 0.75 & \text{Biologia} & 0.43 \\
 \text{Organism} & 0.75 & \text{Soziologie} & 0.28 & \text{Grec ancien} & 0.66 & \text{Organismo vivente} & 0.35 \\
 \text{Philosophy} & 0.72 & \text{Chemie} & 0.28 & \text{Grec} & 0.65 & \text{Tassonomia} & 0.34 \\
 \text{Psychology} & 0.71 & \text{Chemische Verbindung} & 0.27 & \text{Famille de langues} & 0.60 & \text{Ordinamento giuridico} & 0.34 \\
 \text{Behavior} & 0.66 & \text{Reinstoff} & 0.27 & \text{Continent} & 0.57 & \text{Diritto} & 0.31 \\
 \text{Ontology} & 0.52 & \text{Bundesstaat} & 0.26 & \text{Latin} & 0.57 & \text{Homo sapiens} & 0.31 \\
 \text{Entity} & 0.50 & \text{Lebewesen} & 0.22 & \text{Langues italiques} & 0.56 & \text{Stato} & 0.30 \\
 \text{Polity} & 0.50 & \text{Chemische Reaktion} & 0.22 & \text{Notion} & 0.56 & \text{Lingue indoeuropee} & 0.28 \\
 \text{Existence} & 0.50 & \text{Stoffwechsel} & 0.21 & \text{Biologie} & 0.54 & \text{Lingua latina} & 0.26 \\
 \text{State (polity)} & 0.50 & \text{Systematik (Biologie)} & 0.16 & \text{Connaissance} & 0.54 & \text{Lingua (linguistica)} & 0.24 \\
 \text{Physics} & 0.41 & \text{Mathematik} & 0.16 & \text{Liste des {\' E}tats transcontinentaux} & 0.40 & \text{Disciplina (didattica)} & 0.24 \\
 \text{Education} & 0.38 & \text{Astronomie} & 0.14 & \text{Discipline (sp{\' e}cialit{\' e})} & 0.37 & \text{Sport} & 0.21 \\
 \text{Communication} & 0.38 & \text{Kommunikation} & 0.14 & \text{Savoir} & 0.36 & \text{Informazione} & 0.20 \\
 \text{Semiotics} & 0.38 & \text{Tausch (Soziologie)} & 0.14 & \text{France} & 0.30 & \text{XIX secolo} & 0.19 \\
 \text{Meaning-making} & 0.37 & \text{Deutschland} & 0.14 & \text{Syst{\' e}matique} & 0.27 & \text{Prefetto} & 0.19 \\
 \text{Meaning (semiotics)} & 0.37 & \text{Sprache} & 0.14 & \text{Genre (biologie)} & 0.27 & \text{Filosofia} & 0.18 \\
 \text{Learning} & 0.37 & \text{Gesellschaft} & 0.13 & \text{Physique} & 0.26 & \text{Comunicazione} & 0.18 \\
% \text{Country} & 0.34 & \text{Wissen} & 0.13 & \text{Proposition (philosophie)} & 0.25 & \text{Esseri umani} & 0.18 \\
% \text{Language} & 0.34 & \text{Vereinigte Staaten} & 0.13 & \text{Concept (philosophie)} & 0.25 & \text{Mondo} & 0.18 \\
% \text{Politics} & 0.34 & \text{Astronomisches Objekt} & 0.13 & \text{Homo} & 0.24 & %\text{Scienze della Terra} & 0.18 \\ 
 \hline 
  \end{tabular}
  
    \vspace{.5cm}
    
  \begin{tabular}{cc|cc|cc} \hline  
 Russian & \textit{C\textsubscript{B}} & Chinese & \textit{C\textsubscript{B}} &   Japanese &  \textit{C\textsubscript{B}}   \\ \hline
 \text{State} & 1.0 & \text{Organism } & 1.0 & \text{Japan} & 1.0 \\
 \text{Science} & 0.73 & \text{Human } & 0.92 & \text{Europe} & 0.94 \\
 \text{Activity (process)} & 0.69 & \text{Taxonomy } & 0.92 & \text{Eurasian Continent} & 0.85 \\
 \text{Organization} & 0.69 & \text{Biology } & 0.90 & \text{East Asia} & 0.74 \\
 \text{Social group} & 0.66 & \text{Animal} & 0.89 & \text{Eurasia} & 0.73 \\
 \text{Psychology} & 0.65 & \text{Europe} & 0.88 & \text{Earth} & 0.59 \\
 \text{Activity} & 0.64 & \text{Bacteria} & 0.82 & \text{Humanity} & 0.37 \\
 \text{Activity (psychology)} & 0.64 & \text{Gram-negative bacteria} & 0.82 & \text{Human} & 0.35 \\
 \text{Set (mathematics)} & 0.50 & \text{Escherichia coli} & 0.82 & \text{Indo-European Language} & 0.30 \\
 \text{Mathematics} & 0.50 & \text{Developmental biology} & 0.77 & \text{Greek} & 0.28 \\
 \text{Cognition} & 0.37 & \text{Society} & 0.77 & \text{God} & 0.26 \\
 \text{Concept} & 0.33 & \text{Evolutionary developmental biology} & 0.77 & \text{Ancient Greek} & 0.26 \\
 \text{Thinking} & 0.32 & \text{Multicellular organism} & 0.77 & \text{Time} & 0.26 \\
 \text{Matter (Philosophy)} & 0.25 & \text{Cellular differentiation} & 0.77 & \text{Muslim} & 0.26 \\
 \text{Psychic} & 0.24 & \text{Individual} & 0.76 & \text{Quran} & 0.26 \\
 \text{Russia} & 0.23 & \text{History} & 0.75 & \text{Event} & 0.26 \\
 \text{Nature} & 0.19 & \text{Europa} & 0.75 & \text{Country} & 0.24 \\
 \text{Naturalist} & 0.19 & \text{Information} & 0.57 & \text{Time period} & 0.23 \\
 \text{Physics} & 0.19 & \text{Fiber} & 0.57 & \text{Modern times} & 0.23 \\
 \text{Natural history} & 0.19 & \text{Paper} & 0.57 & \text{Federation} & 0.23 \\
% \text{System} & 0.18 & \text{Newspaper} & 0.57 & \text{United States} & 0.23 \\
% \text{Electromagnetic radiation} & 0.16 & \text{News} & 0.57 & \text{State of the United States} & 0.22 \\
% \text{Space (physics)} & 0.16 & \text{Six Ws} & 0.57 & \text{Language} & 0.17   \\ 
\hline 
  \end{tabular}
  }
  \caption{\small {\bf Concepts with highest betweenness centrality \textit{C\textsubscript{B}} for several European and East Asian languages}. The high ranking of Muslim and Quran in Japanese is due to the first link Event $\to$ Quran, which finds itself on the first-link path between United States and God. 
  }\label{tableBCUeurop}
\end{table}

\end{document}